\documentclass[aps,superscriptaddress,twoside,twocolumn,floatfix,a4paper,groupedaddress, pra]{revtex4-1}
\usepackage{color}
\usepackage[usenames,dvipsnames]{xcolor}
\usepackage[unicode=true, breaklinks=false, pdfborder={0 0 1}, backref=false, colorlinks=true, linkcolor=blue, citecolor=blue, urlcolor=blue]{hyperref}
\usepackage{physics}
\usepackage{amsmath,amssymb, amsthm, amsfonts}
\usepackage{dsfont} 
\usepackage{xspace}
\usepackage{xcolor}
\usepackage{changes}
\usepackage{verbatim}
\usepackage{mathtools}
\usepackage{soul}
     \makeatletter
     \renewcommand\@make@capt@title[2]{%
      \@ifx@empty\float@link{\@firstofone}{\expandafter\href\expandafter{\float@link}}%
       {\textbf{#1}}\@caption@fignum@sep#2\quad}%
     \makeatother
\usepackage{braket}
\usepackage{graphicx}
\usepackage{subfigure}
\DeclareMathAlphabet{\mathbbold}{U}{bbold}{m}{n}

\newcommand{\bb}[0]{\begin{eqnarray}}
\newcommand{\ee}[0]{\end{eqnarray}}

\newcommand{\gsim}{\raisebox{-0.13cm}{~\shortstack{$>$ \\[-0.07cm]
      $\sim$}}~}

\begin{document}

\title{Characterizing the performance of heat rectifiers}

\author{Shishir Khandelwal, Mart\'i Perarnau-Llobet, Stella Seah, Nicolas Brunner and G\'eraldine Haack}
\affiliation{Department of Applied Physics, University of Geneva, 1211 Geneva, Switzerland}

\vspace{10pt}

\date{\today}

\begin{abstract}
A physical system connected to two thermal reservoirs at different temperatures is said to act as a heat rectifier when it is able to bias the heat current in a given direction, similarly to an electronic diode. We propose to quantify the performance of a heat rectifier by mapping out the trade-off between heat currents and rectification. By optimizing over the system's parameters, we obtain Pareto fronts, which can be efficiently computed using general coefficients of performance. This approach naturally highlights the fundamental trade-off between heat rectification and conduction, and allows for a meaningful comparison between different devices for heat rectification. We illustrate the practical relevance of these ideas on three minimal models for spin-boson nanoscale rectifiers, i.e., systems consisting of one or two interacting qubits coupled to bosonic reservoirs biased in temperature. Our results demonstrate the superiority of two strongly-interacting qubits for heat rectification. 

\end{abstract}

\maketitle

\section{Introduction}

Managing heat flows in nanoscale devices is a fundamental challenge, notably from the perspective of minimizing heat dissipation and exploiting heat currents for producing electrical power \cite{Giazotto2006, Roberts2011, Li2012, Pekola2015}. A key device in this context is the heat rectifier, first analyzed at the nanoscale in the seminal works ~\cite{Terraneo2002, Li2004, Li2006, Segal2005, Wu2009}. Similarly to an electronic diode, which features a strong resistance in one direction and a low one in the opposite direction, a heat rectifier aims at biasing the flow of heat in a given direction (see also the reviews \cite{Benenti2017, Landi2021}). Remarkably, heat rectification at the nanoscale was successfully demonstrated experimentally over the past decade exploiting different platforms, for example, graphene-based samples~\cite{Wang2017}, circuit QED setups~\cite{Saira2007, Ronzani2018, Senior2020, Gubaydullin2022} and hybrid superconducting-normal devices~\cite{Martinez2015,Strambini2022}.\par

\begin{figure}[t]
    \centering
    \includegraphics[width=1.0\columnwidth]{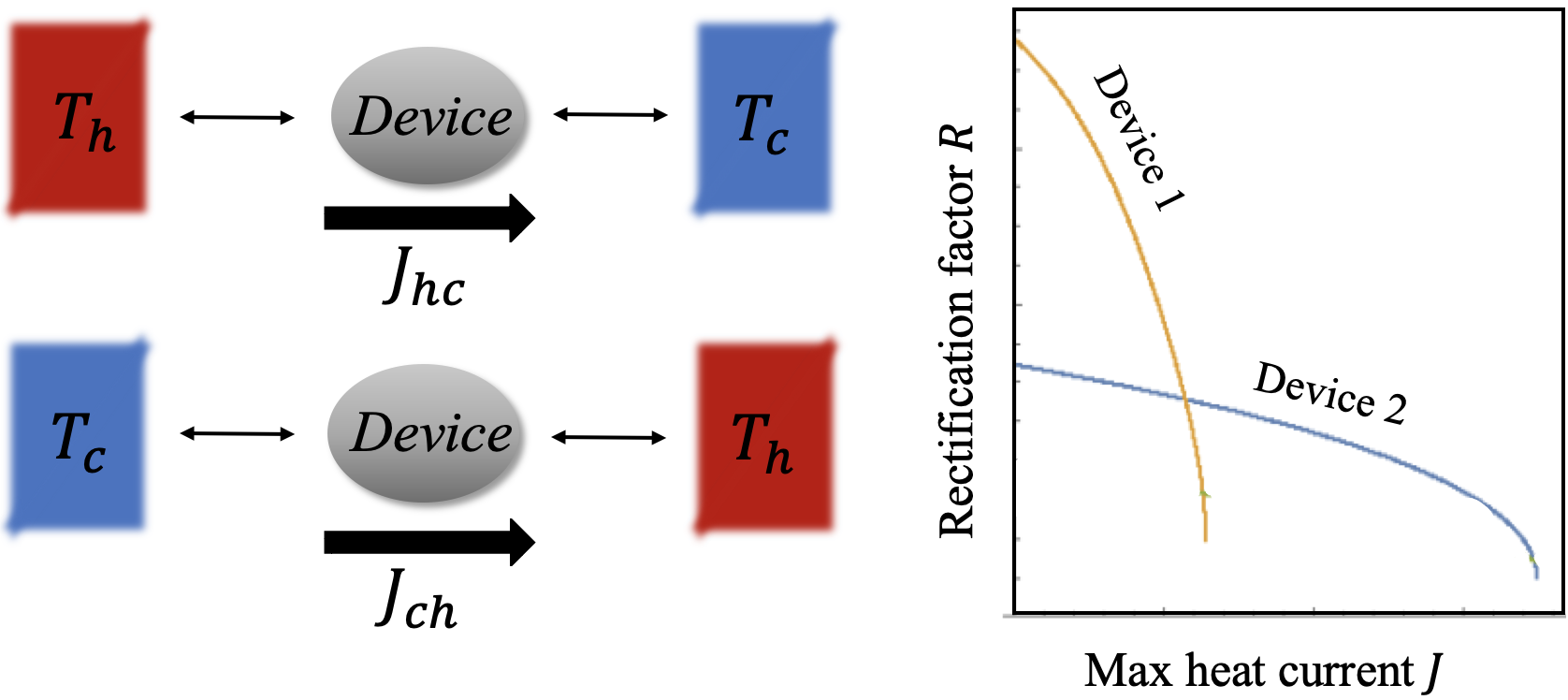}
    \caption{(Left) A device coupled to two thermal reservoirs at different temperatures operates as heat rectifier when it is able to bias the heat current in a given direction, i.e. $ \vert J_{hc} \vert \neq \vert J_{ch} \vert$. This imbalance in the heat currents can be quantified via a rectification factor $R$ defined in Eq.~\eqref{eq:R}. (Right) In order to characterize the performance of a given device as heat rectifier, we propose to plot the rectification factor $R$ versus the maximum heat current $J = \max \{ |J_{ch}|, |J_{hc}| \}$. This allows for comparing the performance of different devices for heat rectification. For example, here device 1 can achieve higher rectification factors than device 2, but only in the regime of small currents. If a larger heat current is required, then device 2 is appropriate.
    }
    \label{fig:scheme}
\end{figure}

The simplest scenario for discussing heat rectification is a two-terminal setup (see Fig.~\ref{fig:scheme}). A (quantum) system is connected to two thermal reservoirs at different temperatures. An ideal rectifier would allow for a strong heat flow when biasing the temperature gradient in one direction, while showing a vanishingly small heat current when biasing the temperatures in the opposite direction. Intuitively, the device must feature a form of asymmetry in order to operate as a heat rectifier. More formally, it was shown that a left-right asymmetry is necessary, as well as non-linearities in the energy spectrum of the device \cite{Terraneo2002,Li2004,Segal2005,Li2006, Wu2009, Wu2009b, Roberts2011, Silva2020, Li2012,Lopez2013,Defaveri2021}. For example, considering a setup with two bosonic thermal reservoirs,  a simple two-level system (qubit) can act as a rectifier, while a harmonic oscillator cannot \cite{Segal2005, Kalantar2021}. For spin (qubit) chains, the simplest way to break left-right symmetry is to consider asymmetric couplings between the system and the left and right reservoirs. Non-linearities in the spectrum can be induced in qubit chains by  means of energy detuning between the constituents of the chain \cite{Man2016,Mascarenhas2016, Joulain2016, Rossell2017, Miranda2017,Tang2018,  Karg2019,Aligia2020,Silva2020, Tesser2021} and/or with non-homogeneous interaction strengths within the chain that can also involve strong coupling \cite{Terraneo2002, Werlang2014, Schuab2016,Sanchez2017, Balachandran2019, Balachandran2019II,Karg2019,Iorio2021, Zhang2021, Upadhyay2021}. This research direction has triggered numerous works in the past years, analyzing platform-specific conditions for achieving heat rectification, for instance in classical and quantum lattices~\cite{Terraneo2002,Li2004,Guimaraes2015,Balachandran2019, Simn2021, RomeroBastida2021, Defaveri2021, Kalantar2021}, investigating the role of quantum interferences, quantum coherence and entanglement \cite{MarcosVicioso2018, Goury2019, Poulsen2022, Palafox2022}, exploiting non-linear quantum circuits \cite{Segal2005, Wu2009,Ruokola2009,Ruokola2011,Marcos-Vicioso2018,Tang2018,Riera-Campeny2019,  Bhandari2021, Tesser2021, Daz2021}, semiconducting devices ~\cite{Fornieri2014, Sanchez2015} and superconducting structures \cite{Fornieri2015, Giazotto2020, Marchegiani2021, Khomchenko2022, Ili2022}.

In order to characterize the performance of a heat rectifier, the standard approach involves defining a rectification factor (see for instance \cite{Segal2005,Wu2009,Mascarenhas2014,Joulain2016,Balachandran2018bb,Landi2021}). While this quantity captures the asymmetry between the two heat currents, it provides in general no information about their magnitudes. For example, it can be the case that a device achieves its maximal rectification factor in the limit of both currents being vanishingly small, as illustrated in Fig. \ref{fig:scheme} with device 1. Clearly, such a regime serves no purpose from a practical perspective. More generally, there appears to be an unavoidable trade-off between heat conduction and rectification in many nanoscale devices (see for example \cite{Mascarenhas2014,Fornieri2014,Mascarenhas2016,Miranda2017,Tang2018,Upadhyay2021} and \cite{Martinez2015} for an experiment), which is reminiscent of the power-efficiency trade-off in heat engines \cite{Benenti2017}. Clearly, the efficiency or optimal operation of a heat rectifier lacks meaning when characterized by the rectification factor alone.\par

In the present work, we investigate the question of characterizing the performance of heat rectifier from an operational and practical perspective. Specifically, we quantify the performance of a given device by mapping out the trade-off between heat rectification $R$ and the corresponding largest heat current $J$ (refer to the next section for precise definitions). This can be visualized on a 2D plot ($R$ vs $J$). By optimizing over the system's parameters, one obtains Pareto fronts, which characterize the device's performance as heat rectifier. We show that these Pareto fronts can be efficiently computed using a general form of coefficients of performance (COP). \par 

This method allows one to address a number of questions that arise naturally in practice. For example, given a minimal desired heat current $J$, what is the maximal rectification factor $R$ that can be achieved for a given device? Importantly, it also allows for a fair comparison between different devices.  We illustrate the relevance of these ideas using three minimal models for rectifiers sketched in Fig.~\ref{fig:plot1}. These consist of a system of one or two interacting qubits coupled to bosonic reservoirs, inspired by the spin-boson rectifier first introduced in Ref.~\cite{Segal2005}. Each device can be operated as heat rectifier by tuning a minimal number of settings, including left-right asymmetry of the couplings to the reservoirs, energy detuning of the qubits and strong inter-qubit interaction. These models are paradigmatic for quantum technologies, and the simplest spin-boson rectifier has recently been realized experimentally in superconducting platforms \cite{Ronzani2018, Senior2020}. Using master equations, we analyze the heat currents and rectification factors in the different devices and provide microscopic insights into the mechanisms involved in heat rectification at the quantum scale. For each device, we perform numerical optimization of the COP over the device's parameters, which allows us to compare their performances via Pareto fronts. We conclude with a number of open questions.

\section{Characterizing the performance of a heat rectifier}
\label{sec:characterize}

We consider the setup depicted in Fig.~\ref{fig:scheme}. A device (system) is coupled to two thermal reservoirs, referred to as left and right, with temperatures $T_L$ and $T_R$ respectively. We are interested in the situation when the two reservoirs are at different temperatures and the same chemical potentials, which places the system in a regime out of thermal equilibrium. Our focus here is on the steady state, in which a constant heat current flows through the device, from the hot reservoir to the cold one. The main task is to investigate the difference in heat currents when biasing the temperatures of the two reservoirs in one direction or the other. So we are interested in the two heat currents
\begin{align}
 J_{hc} &\coloneqq J(T_L = T_h, T_R = T_c)   \\
 J_{ch} &\coloneqq  J(T_L = T_c, T_R = T_h) \,.\nonumber
\end{align}
Note that here heat currents always flow from hot to cold, and our main concern is therefore the intensity of these currents. Heat rectification is said to occur whenever these two currents are unequal in magnitude, i.e.,
\begin{align}
\vert J_{hc} \vert \neq \vert J_{ch} \vert\,.
\end{align}In order to capture the magnitude of this effect, it is conventional to introduce a rectification factor defined as \cite{Roberts2011,Landi2021}
\begin{align}
\label{eq:R}
R = \left\vert \frac{J_{hc} + J_{ch}}{J_{hc} - J_{ch}} \right\vert\,.
\end{align}When the two currents match exactly in intensity, i.e. $J_{hc} = - J_{ch}$ (heat flows from hot to cold, hence the minus sign when reversing the temperatures), we have that $R=0$, and there is no rectification. When there is an imbalance between the two heat currents, we have that $R>0$ and the device acts as a rectifier. An ideal rectifier would have $R=1$, meaning that heat can only flow in one direction. \par 
At this point however, it is crucial to realize that the above rectification factor $R$ captures only the relative intensity of the two heat currents, independently of their actual intensities. Therefore, two devices can have the same rectification factor $R$, while the respective heat currents may have several orders of magnitude difference. Hence from an operational point of view, $R$ provides only an incomplete description of the rectifier, and from a practical perspective, more information is needed to assess the performance of a device.\par 
This motivates us to present a more general method for characterizing the performance of a heat rectifier, in order to properly account for the trade-off between heat rectification and conduction. Specifically, we define another quantity of interest,
\begin{align}
\label{eq:maxJ}
J = \max \{ |J_{hc} |, |J_{ch}| \}\,,
\end{align}which represents the largest of the two currents. Then, we investigate on a 2D plot the behaviour of the rectification factor $R$ as a function of the maximum current $J$. By optimizing over the relevant parameters of a system, one can obtain so-called Pareto fronts (see Fig.~\ref{fig:scheme}) which then quantify the trade-off between heat rectification and conduction. In practice, this allows one to understand what level of rectification can be expected, under the constraint that the heat current $J$ is not below a certain threshold value. Moreover, it allows one to meaningfully compare the performance of two devices as heat rectifiers, taking into account difference different regimes of operation. For example, one device may achieve higher rectification in the regime of low heat currents, while another device may be preferable when larger currents are required, as illustrated in Fig.~\ref{fig:scheme}.\par 

In order to efficiently compute the Pareto fronts, it is useful to consider the following family of coefficients of performance (COP) $\{\eta_\alpha\}_\alpha$, where
\bb
\label{eq:eta}
\eta_\alpha = \alpha \, R + (1 - \alpha) \, J\,,
\ee
with $\alpha \in [0,1]$ is a parameter that captures the relative weight attached to rectification $R$ and maximal heat current $J$. For given value of $\alpha$, one can maximize $\eta_\alpha$ over the system's parameters. By repeating the procedure for different values of $\alpha$, one can then draw the desired Pareto fronts capturing the trade-off between $R$ and $J$ and hence giving a full description of the performance of the device as heat rectifier.\par 
In the following, we will illustrate the relevance of these ideas considering three minimal designs for quantum heat rectifiers. Importantly, this will allow us to compare their respective performances and draw some general conclusions. More generally, these methods can of course be readily applied to any device, allowing for the comparison of very different types of heat rectifiers. Finally, we note that our approach can also be readily adapted to the use of alternative definitions for the rectification factor, which have been used in the literature (see for example, \cite{Landi2021,Segal2005,Tang2018,Riera-Campeny2019,Miranda2017}).

\section{Minimal models for spin-boson heat rectifiers}
\label{sec:model}

\begin{figure}[t]
    \centering
    \includegraphics[width=1.0\columnwidth]{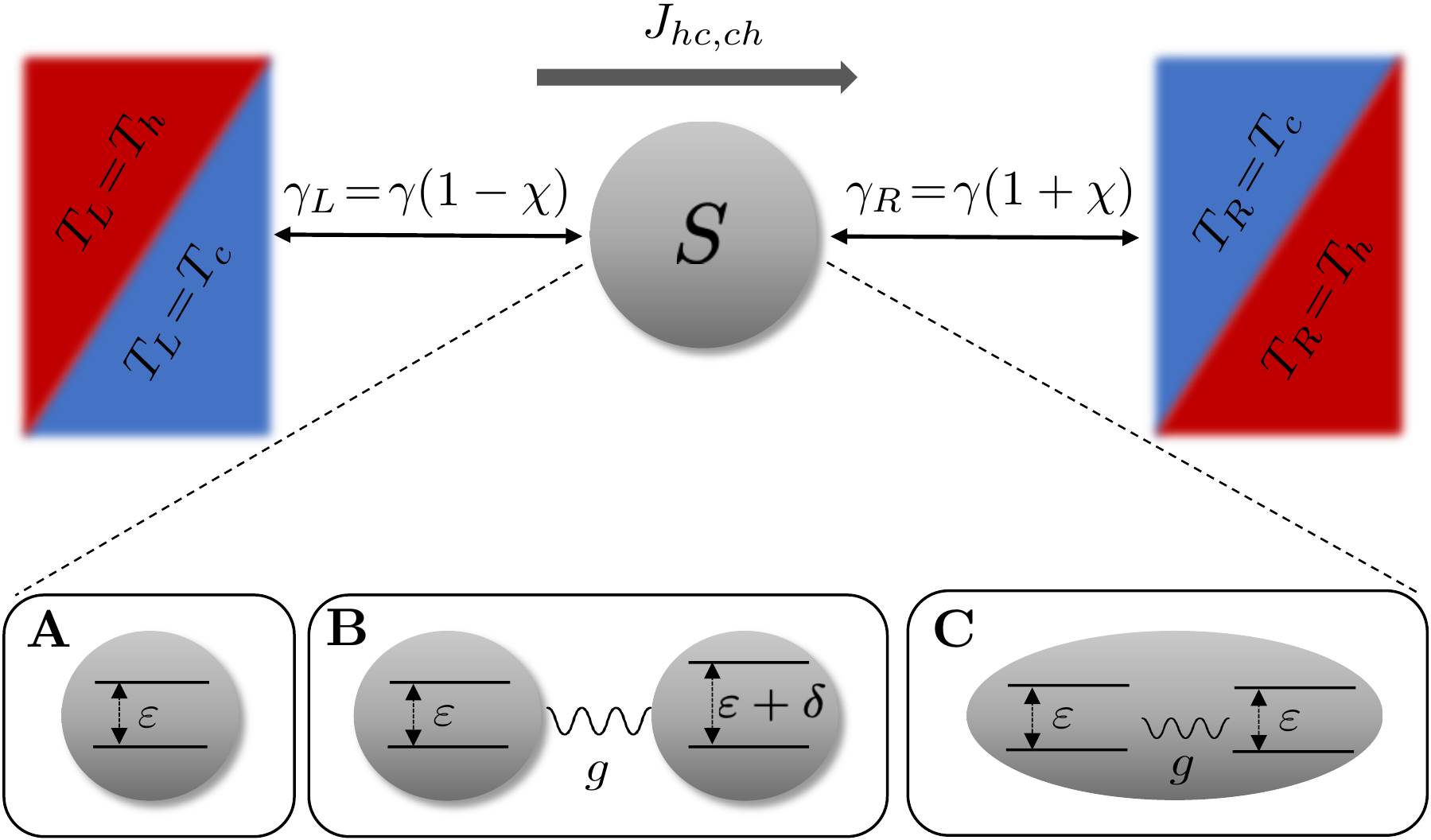}
    \caption{Three minimal spin-boson devices \textbf{S = A, B, C} made of one or two interacting qubits, coupled to two thermal reservoirs with tunable coupling rates characterized by the asymmetry parameter $\chi$. Temperature gradient can be reversed in order to investigate heat rectification properties of the different devices; left and right reservoirs can be set to hot and cold temperatures, or vice-versa. The three minimal models for achieving heat rectification are characterized by their key parameters. \textbf{A} - a single qubit with energy transition $\varepsilon$, \textbf{B} - two weakly-coupled qubits with a possible energy detuning $\delta$ and \textbf{C} - two strongly-coupled qubits characterized by the inter-qubit coupling, $g$. 
    }
    \label{fig:plot1}
\end{figure}In this work, we consider minimal models for quantum heat rectifiers, in which a system of one or two qubits is connected to two thermal reservoirs, as shown in Fig.~\ref{fig:plot1}. We consider three specific models, each of which can achieve heat rectification by tuning different control parameters. Specifically, we have: 

\begin{itemize}
    \item Device \textbf{A}: spin-boson system, featuring a single qubit with energy gap $\varepsilon$.
    
    \item Device \textbf{B}: two weakly interacting qubits, detuned in energy by $\delta$, i.e. with the energy gaps $\varepsilon$ and $\varepsilon + \delta$.
    
    \item Device \textbf{C}: two strongly interacting qubits degenerate in energy $\varepsilon$, with inter-qubit coupling $g$.

\end{itemize}

For each device, one can tune independently the couplings to the reservoirs, hence breaking the left-right symmetry. Denoting the couplings $\gamma_L, \gamma_R$ with the left and right reservoirs, the asymmetry can be captured by a single parameter $\chi\in\left[-1,1\right]$, defined by setting $\gamma_L = \gamma (1- \chi)$ and $\gamma_R = \gamma (1+ \chi)$. Here $\gamma$ is a bare coupling rate set by microscopic parameters. The regime $\chi=0$ corresponds to symmetric couplings $\gamma_L = \gamma_R = \gamma$ and $\chi=\pm1$ implies that one of the couplings is zero, i.e., the system is only coupled to a single reservoir and can no longer function as a rectifier. \par

We evaluate the heat currents in the stationary state within a master equation framework. We assume Markovian bosonic reservoirs and weak coupling between the reservoirs and devices. Hence the evolution equation for the reduced density operator $\rho^{(i)}$ of the device $i= A,B,C$ can be expressed in a Lindblad form \cite{Lindblad1976,Gorini1976,Breuer2007} (we set $\hbar = k_B = 1$ throughout the manuscript),
\bb
\label{eq:ME}
&&\dot{\rho}^{(i)} =  -i [H^{(i)}, \rho^{(i)}]  + \nonumber \\
&&\sum_{\sigma=L,R} \sum_k  \Big( \gamma_\sigma^+(E_{k,\sigma}^{(i)}) \mathcal{D}[A_{k,\sigma}^{(i)}] \rho^{(i)} + \gamma_\sigma^-(E_{k, \sigma}^{(i)}) \mathcal{D}[A^{(i) \dagger}_{k,\sigma}] \rho^{(i)} \Big)\,. \nonumber \\
&&
\ee
The Lindblad evolution equation is composed of a unitary part through the commutator involving the Hamiltonian $H^{(i)}$ of device $i$  $(i=\text{\textbf{A}, \textbf{B}, \textbf{C}})$ and a dissipative part due to the presence of the left and right reservoirs labelled with the index $\sigma = L, R$. Dissipation occurs through the superoperators $\mathcal{D}$ acting onto $\rho^{(i)}$ defined as,
\bb
\mathcal{D}[X]\rho := X \rho X^\dagger - \frac{1}{2} \left\{ X^\dagger X, \rho \right\} \,,
\ee
where $\{\cdot , \cdot \}$ is the anticommutator. The jump operators $A_{k,\sigma}^{(i)}$ and $A^{(i) \dagger}_{k,\sigma}$ respectively describe a single excitation jumping \textit{out} of the system or a single excitation jumping \textit{in} to the system 
to and from reservoir $\sigma = L,R$ at energy $E_k$. This energy $E_k$ corresponds to the energy gap between the two states involved in the jump. 
The corresponding outgoing and incoming rates are  $\gamma_\sigma^-(E_k^{(i)})$ and $\gamma_\sigma^+(E_k^{(i)})$. They are set by a bare coupling rate $\gamma$ introduced above, and by the quantum statistics of the reservoirs. Importantly, they are evaluated at the energy transition $E_k^{(i)}$ and at the temperature $T_\sigma$ of the reservoir $\sigma=L,R$. As we consider here exclusively bosonic reservoirs, the rates take the form \cite{Breuer2007},
\begin{equation}
\begin{aligned}
\label{eq:rates}
&\gamma_\sigma^-(E_k^{(i)}) &\equiv& \gamma_\sigma \, \big( 1 + n_B (E_k^{(i)}, T_\sigma)\big) \\
&\gamma_\sigma^+(E_k^{(i)}) &\equiv& \gamma_\sigma \,\,  n_B (E_k^{(i)}, T_\sigma),
\end{aligned}
\end{equation}
where $n_B$ is the Bose-Einstein distribution $n_B (E, T) = (e^{E/T} - 1)^{-1}$. The steady state $\rho^{(i)}_{ss}$ of device i is found by imposing $\dot{\rho}^{(i)} = 0$, and is then used to derive the steady-state (stationary) heat current in the right reservoir for device $i$. In a two-terminal device with charge conservation, it is defined as \cite{Benenti2017,Landi2021},
\begin{align}
\label{eq:current}
J^{(i)} = \dot{Q}^{(i)}_R - \dot{Q}^{(i)}_L,
\end{align}
with
\begin{widetext}
\bb
\dot{Q}^{(i)}_\sigma = \text{Tr} \left\{ H^{(i)} \sum_k  \Big( \gamma_\sigma^+(E_k^{(i)}) \mathcal{D}[A_{k,\sigma}^{(i)}] \rho^{(i)}_{ss} + \gamma_\sigma^-(E_k^{(i)}) \mathcal{D}[A^{(i) \dagger}_{k,\sigma}] \rho^{(i)}_{ss} \Big) \right\}\,, \quad \sigma = L,R \,.
\ee
\end{widetext}
In the following, we present analytical results for the three devices \textbf{A, B, C} sketched in Fig.~\ref{fig:plot1}. We then illustrate the unavoidable trade-off between heat conduction and heat rectification and compare their performance using Pareto fronts computed by optimizing the family of COPs $\eta_\alpha$.

\section{Device \textbf{A} : Single-qubit rectifier}
\label{sec:devA}

\begin{figure}[t]
\centering
\includegraphics[width=\columnwidth]{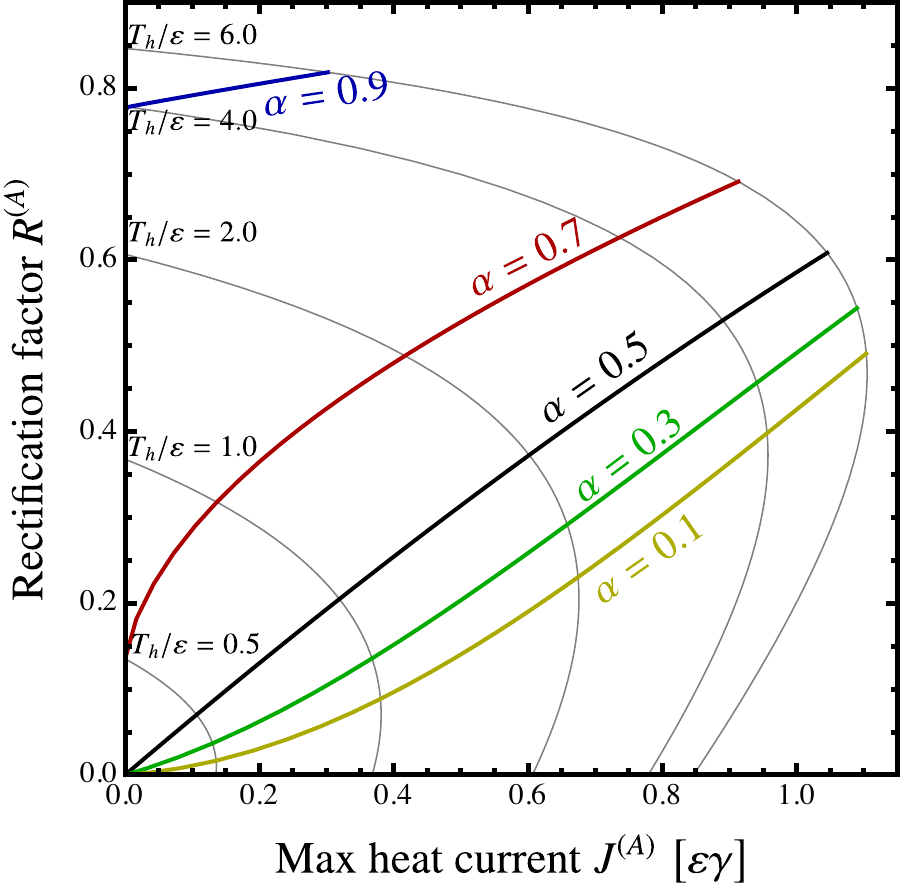}
\caption{Heat conduction versus heat rectification for device \textbf{A} and optimal operation regime according to COP $\eta_\alpha$. The gray lines are parametric plots of the maximum heat current, $J^{(A)}=\text{max}\left\{\left\lvert J_{hc}^{(A)}\right\rvert,\left\lvert J^{(A)}_{ch}\right\rvert \right\}$ and $R^{\left( A\right)}$ with $0\leq\chi\leq 1$ as the varying parameter. For hot temperatures between $0<T_h/\varepsilon\leq 6$, $\eta_\alpha$ is optimized over the the asymmetry $\chi$ for various values of the convex parameter $\alpha$. The cold temperature is fixed at $T_c/\varepsilon=0.01$. These curves represent the corresponding optimal regimes of operation of the rectifier.} 
\label{fig:sing}
\end{figure}
Device \textbf{A} consists of a single qubit with energy gap $\varepsilon$ described by the Hamiltonian,
\bb
H^{(A)} = \varepsilon \, \sigma_+\sigma_-,
\ee
with $\sigma_+$ and $\sigma_-$ the raising and lowering operators in the computational basis of the qubit $\{ \ket{1}, \ket{0}\}$, $\sigma_- \ket{1} = \ket{0}$ and $\sigma_+ \ket{0} = \ket{1}$. The jump operators entering the master equation Eq.~\eqref{eq:ME} are raising and lowering operators, $\sigma_-, \sigma_+$ operators, capturing single particle tunneling between system and reservoirs. For device \textbf{A}, the computational basis is the energy eigenbasis and the jump rates are therefore evaluated at the unique energy scale of the system, $\varepsilon$. The steady state for this simple system reads, 
\begin{align}
    \rho_{\text{ss}}^{\left\langle A\right\rangle}=\frac{1}{ \Gamma }\left(
\begin{array}{cc}
 { \Gamma^+} & 0 \\
 0 & { \Gamma^-} \\
\end{array}
\right),
\end{align}where $\Gamma^\pm\coloneqq \gamma_h^\pm + \gamma_c^\pm $ and $\Gamma\coloneqq \Gamma^- + \Gamma^+$. 
\par The current in the right reservoir can be computed using Eq.~\eqref{eq:current}. In the presence of asymmetric couplings, $\gamma_L \neq \gamma_R$, this system can operate as a heat rectifier~\cite{Segal2005} and has been recently experimentally realized on a circuit QED platform \cite{Senior2020}. The currents $J_{hc}$ and $J_{ch}$ can be expressed in the compact form,
\begin{equation}
\begin{aligned}
\label{eq:Jsing}
J^{(A)}_{hc/ch} &= \frac{ 2\, \varepsilon \, \gamma \,  \left(1-\chi^2\right) }{1+\Sigma + \chi \, \Delta_{hc/ch} } \, \Delta_{hc/ch} \\
&= \frac{\pm \, 2\, \varepsilon \, \gamma \,  \left(1-\chi^2\right) }{1+\Sigma \mp \chi\, \Delta_{hc} } \, \Delta_{hc},
\end{aligned}
\end{equation}
with the notations
\bb
&&\Delta_{hc} = n_B (\varepsilon, T_h) - n_B (\varepsilon, T_c) \label{eq:delta}\,, \\
&& \Sigma =  n_B (\varepsilon, T_h) + n_B (\varepsilon, T_c)\,.\nonumber \label{eq:sigma}
\ee
We evidently have for a single qubit characterized by $\varepsilon$ only, that $\Delta_{ch} = - \Delta_{hc}$, hence the form of the heat currents in Eq. \eqref{eq:Jsing}. As expected from quantum transport theory, the current is proportional to the difference of the distributions of the two reservoirs. For reservoirs with equal chemical potentials, if $T_L = T_R$, this difference vanishes for a single qubit and no transport occurs. One can also note that in absence of left-right asymmetry, $\chi = 0$, the currents are the same upon exchanging the temperatures of the two reservoirs, $J_{hc}^{(A)} = J_{ch}^{(A)}$, forbidding any rectification effect. This is also evidenced by inserting Eq.~\eqref{eq:Jsing} into Eq.~\eqref{eq:R} to obtain the rectification factor for device $\textbf{A}$,
\begin{align}
\label{eq:rsing}
R^{(A)}=  \frac{\left|\chi\right|}{1+\Sigma} \, \Delta_{hc}\,.
\end{align}
Again, in the absence of thermal bias $\Delta_{hc} = 0$, no current and hence no rectification are present. $R^{(A)}$ depends linearly on the asymmetry factor $\chi$ and the maximum value of $R^{(A)}$ occurs for maximum asymmetry,~$\chi\to\pm1$, which corresponds to a vanishing $J^{(A)} = \text{max}\{ J^{(A)}_{hc}, J^{(A)}_{ch}\}$, independently of the coupling rate $\gamma$. We note that this linear dependence reflects the same physics as discussed in \cite{Segal2005} with a different definition of the rectification factor. Since $\Delta_{hc}/(1+\Sigma)$ is upper bounded by 1, Eq.~\eqref{eq:rsing} implies that the rectification factor is upper bounded by $\lvert \chi\rvert$. Such an upper bound was discussed earlier in \cite{Tesser2021} in the context of quantum dots coupled with fermionic reservoirs. \par  
In Fig.~\ref{fig:sing}, we show the behaviours of $R^{(A)}$ versus $J^{(A)}$ (defined in Eq.~\eqref{eq:maxJ}) as a function of the asymmetry parameter $\chi$ for fixed temperatures of the cold reservoir ($T_c = 0.01\, \varepsilon$) and different temperatures of the hot reservoir $T_h \,[\varepsilon] = 0.5, 1 , 2 ,4 ,6$ (gray thin lines). These curves clearly show the trade-off between heat conduction and heat rectification. Maximal rectification comes at the cost of zero heat current. Utilizing the COP $\eta_\alpha$, we determine optimal values of $\chi$ to favor heat conduction or heat rectification, by varying the parameter $\alpha$. This is shown by the thick colored solid lines corresponding to different values of $\alpha$ in Eq.~\eqref{eq:eta}. For each temperature bias ($T_h$ is varied a $T_c$ being fixed), optimal values of $\chi$ are found numerically. As more weight is put on heat rectification, heat conduction decreases, and vice-versa. For instance, for $\alpha = 0.9$, rectification is close to its maximum $R^{(A)} \sim 0.8$, whereas heat current remains around $0.4$ (max values for heat currents reach 1.1 in arbitrary units). Again, this reflects the trade-off between the two properties present in nanoscale devices. \par
Finally, it is worth discussing the regime without rectification, i.e. $R=0$. In this case, the device behaves symmetrically in terms of the two heat currents, and we have that $J_{hc}=-J_{ch}$. Here, this requires perfectly symmetrical couplings to the reservoirs, i.e. $\chi=0$, since the device is otherwise perfectly symmetrical. Also, we note that this regime is in fact rather rare, as it requires a precise tuning of the parameters.\\ \\



\begin{figure*}
\centering
\begin{subfigure}
  \centering
  \includegraphics[width=.477\linewidth]{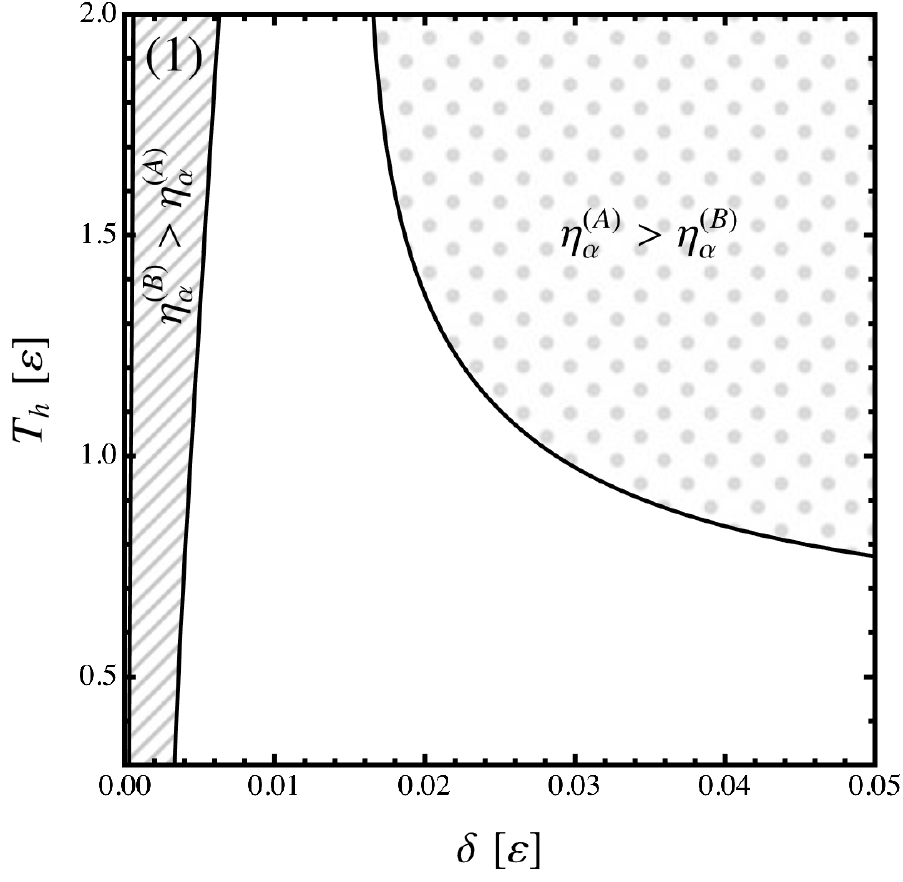}
\end{subfigure}%
\begin{subfigure}
  \centering
  \includegraphics[width=.46\linewidth]{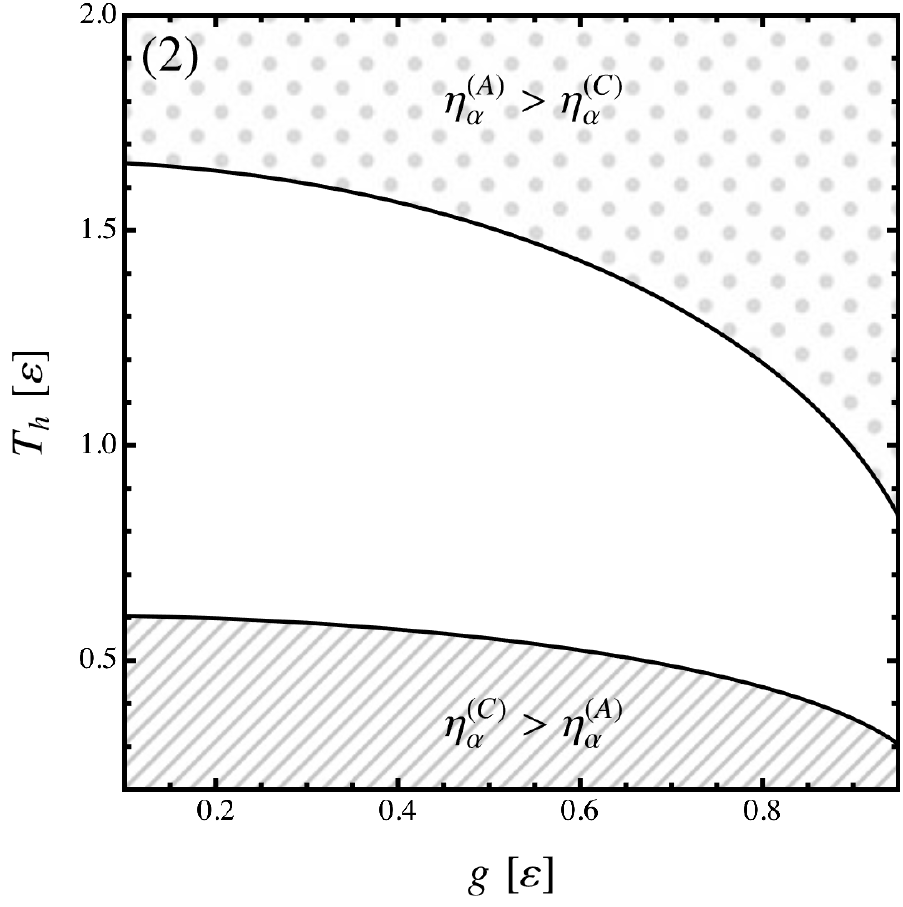}
\end{subfigure}
\caption{Panel (1): Region plot in the parameter space of $T_h$ and $\delta$,  
comparing heat currents and rectification factors of device \textbf{A} (single qubit) and device \textbf{B} (two weakly-interacting qubits). Other parameters: $\varepsilon=1$, $g/\varepsilon=0.05$, $\gamma/\varepsilon=0.001$, $\chi=0.4$ and $T_c/\varepsilon=0.01$. Panel (2): Region plot in the parameter space of $T_h$ and $g$,  
comparing heat currents and rectification factors of device \textbf{A} (single qubit) and device \textbf{C} (two strongly interacting qubits). Other parameters: $\varepsilon=1$, $\gamma/\varepsilon=0.001$, $\chi=0.4$ and $T_c/\varepsilon=0.01$. The performance of the rectifiers can be unambiguously compared without using $\eta_\alpha$ in the dotted and dashed regions, while it importantly, depends on $\alpha$ in the white region.}
\label{fig:regionAB}
\end{figure*}

\section{Device \textbf{B}: \\ Two weakly-interacting qubits}
\label{sec:devB}
Device \textbf{B} consists of two weakly-interacting qubits (left and right) with bare energies $\varepsilon_L$ and $\varepsilon_R$. The interaction between the two qubits is of flip-flop type, with strength set by $g$ (see Fig.~\ref{fig:plot1}). The weak interaction regime is defined by the relative strengths $g$ versus the bare coupling to the reservoirs $\gamma$, $g \leq \gamma \ll \varepsilon_L, \varepsilon_R$ and the total Hamiltonian of device is given by,
\bb
&&H^{(B)} = \varepsilon_L\sigma^{(L)}_+\sigma^{(L)}_- + \varepsilon_R \sigma^{(R)}_+\sigma^{(R)}_- + g (\sigma_+^{(L)}\sigma_-^{(R)}+\sigma_-^{(L)}\sigma_+^{(R)})\,. \nonumber \\
&&
\ee
In the weak-interaction regime, it is relevant to work in the computational basis of the two qubits $\{\ket{11}, \ket{10}, \ket{01}, \ket{00}\}$. The jump operators in Eq.~\eqref{eq:ME} are local and are given by \cite{Hofer2017, Gonzalez2017, Mitchison2018, Khandelwal2020, Landi2021},
\bb
A_{k,\sigma}^{(B)} &=& \sigma_-^{\sigma} \\
A_{k,\sigma}^{(B) \dagger} &=& \sigma_+^{\sigma}\,.\nonumber
\ee
There is only a single relevant energy per qubit, $E_{k,L}^{(i)} = \varepsilon_L$ and $E_{k,R}^{(i)} =\varepsilon_R$. The steady-state solution to the corresponding  master equation can be found in \cite{Khandelwal2020}. Note that this device has also been considered in Ref.~\cite{Miranda2017} in the context of heat rectification.\par

When the two qubits are degenerate in energy, $\varepsilon_L = \varepsilon_R = \varepsilon$, the transition energies involved in the jumps are all equal to $\varepsilon$ and the steady state takes a simple form. The heat currents are calculated according to Eq.~\eqref{eq:current}, and can be expressed exactly in terms of the heat current of device \textbf{A} as follows,
\bb
\label{eq:JBdeg}
J_{hc/ch}^{(B), \text{deg}} = \frac{1}{1+\Gamma_L\Gamma_R/4g^2} \, J_{hc/ch}^{(A)}\,.
\ee
with $\Gamma_L \coloneqq \gamma_L^+ + \gamma_L^-$ and $\Gamma_R \coloneqq \gamma_R^+ + \gamma_R^-$, for convenience. Furthermore, the rectification factor for this case is exactly equal to $R^{(A)}$,
\bb
\label{eq:RBdeg}
 R^{(B),\text{deg}}=R^{(A)}.
\ee
Since $1/(1+\Gamma_L\Gamma_R/4g^2) \leq 1$, Eqs.~\eqref{eq:JBdeg} and \eqref{eq:RBdeg} imply that the heat conduction ability of device \textbf{B} for energy degenerate qubits is always worse than one of device \textbf{A} and that heat rectification properties are the same. Equation~\eqref{eq:RBdeg} can be intuitively understood as the proportionality factor between $J_{hc/ch}^{(B),\text{deg}}$ and $J_{hc/ch}^{(A)}$ is symmetric upon exchanging the temperatures of the left and right reservoirs. Hence, in this particular case, one can directly see 
that device \textbf{A} always performs
better than device \textbf{B} with degenerate qubits. Let us remark that this statement also holds for a chain of $N$ energy-degenerate qubits as the heat current turns out to be independent of the length of the chain, see for example, Ref.~\cite{Landi2021}.\par

The situation is  different when the two weakly interacting qubits of device \textbf{B} are non degenerate in energy,   $\varepsilon_L \neq \varepsilon_R$, i.e., when there is an energy detuning $\delta$ between both. In the following, we define the energy of the left qubit to be the reference $\varepsilon_L = \varepsilon$, and we define the energy of the right qubit as $\varepsilon_R =  \varepsilon+\delta$ as sketched on Fig.~\ref{fig:plot1}. The master equation with local jump operators remains valid in the presence of detuning when all energy scales are much smaller than the bare energies of the qubits. Hence, here we must further impose that $\delta\ll\varepsilon_L, \varepsilon_R$. In Ref.~\cite{Khandelwal2020}, the heat current was derived analytically, and for relatively large detuning was shown to be smaller in magnitude compared to device \textbf{B} with energy-degenerate qubits. However, for smaller detuning, when investigating rectification properties, device \textbf{B} turns to be advantageous over device \textbf{A} in some parameter's range. This is illustrated in Fig.~\ref{fig:regionAB}, where we compare the COPs of devices \textbf{A} and \textbf{B} as functions
of the hot temperature $T_h$ and detuning $\delta$. The shaded region corresponds to the parametric region in which for \textit{any value} of $\alpha$ in Eq.~\eqref{eq:eta}, $\eta_\alpha^{(B)}>\eta_{\alpha}^{(A)}$,  because,

\begin{equation}
\begin{aligned}
 J^{(B)} > J^{(A)}, \quad \text{and} \quad R^{(B)} > R^{(A)} \,.
\end{aligned}
\end{equation}In other words, in the shaded region, independently of the weight put on heat conduction or heat rectification, device \textbf{B} performs better than device \textbf{A}. In contrast, the dotted region corresponds to the opposite situation, where device \textbf{B} performs worse than device \textbf{A} for any value of $\alpha$. In the white region, the possible advantage of one device over another depends on the value of $\alpha$. We note that any advantage of device \textbf{B} over device \textbf{A} vanishes in the limit $g\to0$, since the heat current in device \textbf{B} vanishes in this limit. \par

As we did for device \textbf{A}, we mention here the conditions for $R=0$ (i.e. symmetric heat currents $J_{ch}=-J_{hc}$) in device \textbf{B}. When the two qubits are degenerate in energy, then the symmetric $R=0$ regime requires symmetric couplings to the reservoirs ($\chi=0$). Interestingly, it is possible to reach the symmetric $R=0$ regime even when the couplings to the reservoirs are different ($\chi\neq0$), by taking energy-detuned qubits. Detuning adds an additional asymmetry, and can be carefully adjusted to offset the asymmetry due to couplings; the exact condition can be computed using the results in \cite{Khandelwal2020}.

\section{Device \textbf{C}:~\\~Two strongly interacting qubits}
\label{sec:devC}

We now consider two strongly interacting qubits, degenerate in energy. This regime is characterized by the energy scales, $\varepsilon\gsim g \gg\gamma_L,\gamma_R$ and a global master equation has been shown to capture adequately the dynamics of the device in this regime \cite{Hofer2017,Khandelwal2020, Landi2021}. This global master equation corresponds to Eq.~\eqref{eq:ME}, with jump operators acting onto the energy eigenstates of the two interacting qubits, $\{ \ket{11},\, \ket{\varepsilon_+},\, \ket{\varepsilon_-},\,   \ket{00}\}$ with respective eigenenergies $2 \varepsilon,\, \varepsilon_\pm = \varepsilon \pm g$ and $0$. The transition rates in Eq.~\eqref{eq:rates} are evaluated at energy transitions between the energy eigenstates involved in the jumps. We again refer to Ref.~\cite{Khandelwal2020} for explicit expressions in this regime. Using the steady-state solution for the reduced density operator, the heat current follows from Eq.~\eqref{eq:current}.
\begin{widetext}
\bb
\label{eq:JC}
   J^{(C)}_{hc/ch} =\gamma\left(1-\chi^2\right)\left(   \frac{\pm\varepsilon_- \, \Delta_{hc}(\varepsilon_- )}{1+\Sigma (\varepsilon_-) \mp \chi\Delta_{hc}(\varepsilon_-)}  + \frac{\pm\varepsilon_+ \,\Delta_{hc}(\varepsilon_+ )}{1+\Sigma (\varepsilon_+) \mp \chi\Delta_{hc} (\varepsilon_+)} \right),
\ee
\end{widetext}
with $\Delta_{hc}$ and $\Sigma$ defined earlier by Eqs.~\eqref{eq:delta} and \eqref{eq:sigma} respectively. The energy argument of these functions indicates the energy transitions at which the Bose-Einstein distributions in $\Delta_{hc}$ and $\Sigma$ should be evaluated.

The above form of the heat current bears a striking similarity with the currents of devices \textbf{A} and \textbf{B} as each term in the sum exhibits the same structure. The sum of two terms evaluated at $\varepsilon_-$ and $\varepsilon_+$ in Eq.~\eqref{eq:JC}  reflects how strong coupling enriches transport by involving more states (here energy eigenstates $\ket{\varepsilon_+}$ and $\ket{\varepsilon_-}$).  As expected, each term is proportional to the difference in Bose-Einstein distributions $\Delta_{hc}$. Compared to device \textbf{A}, the Bose-Einstein distributions are now evaluated at the eigenenergies $\varepsilon_\pm$. The rectification factor $R^{(C)}$ takes the following form,
\bb
\label{eq:RC}
    R^{(C)} = \left\lvert \frac{\varepsilon_-\mathcal P\left( \varepsilon_-\right)+ \varepsilon_+\mathcal P\left( \varepsilon_+\right)}{\varepsilon_-\mathcal Q\left( \varepsilon_-\right)+ \varepsilon_+\mathcal Q\left( \varepsilon_+\right)}  \right\rvert,
\ee
with the definitions of terms appearing in the numerator,
\bb
    \mathcal P\left( \varepsilon_\pm \right) \coloneqq \frac{\chi \, \Delta_{hc}(\varepsilon_{\pm})^2}{\left( 1+\Sigma(\varepsilon_{\pm}) \right)^2 -\chi^2\,\Delta_{hc}(\varepsilon_{\pm})^2}
\ee
and in the denominator,
\bb
\mathcal Q\left( \varepsilon_\pm \right) \coloneqq \frac{\Delta_{hc}(\varepsilon_{\pm})\left(  1+ \Sigma(\varepsilon_{\pm}) \right)}{\left( 1+\Sigma(\varepsilon_{\pm}) \right)^2 -\chi^2\, \Delta_{hc}(\varepsilon_{\pm})^2}\,.
\ee

As expected, Eq.~\eqref{eq:RC} shows that there is no rectification in absence of asymmetry in the coupling rates, $\chi=0$. We note that this is the only regime in which there in no rectification shown by the device; the reason is that the device is internally symmetrical. In contrast, the heat current falls to zero at maximum asymmetry $\chi=\pm1$, see factor $1-\chi^2$ in Eq.~\eqref{eq:JC}, reflecting again the trade-off between heat conduction and heat rectification in this device. Let us note that the trade-off here is a bit more subtle than in device \textbf{A}, as $R^{(C)}$ is no longer linear in $\chi$; its maximum does not necessarily lie at $\chi=\pm 1$. In Fig.~\ref{fig:regionAB}, we show how the advantage of device \textbf{C} can be discussed with respect to device \textbf{A} using $\eta_\alpha$, similar to the discussion between devices \textbf{A} and \textbf{B}. Based on the COP $\eta_\alpha$, it is possible to determine two regions in the parameter space where one device is unambiguously better than the other one. Then, in the region left in blank, which device performs better depends on the convex parameter $\alpha$. For a given set of hot temperature $T_h$ and coupling strength $g$, device \textbf{C} may be operated as a better heat conductor or a better heat rectifier than device \textbf{A}.\par

Let us note that the advantage of device \textbf{C} over device \textbf{A} is expected to decrease when $g \rightarrow 0$. This can be understood analytically. In the limit $g\to0$, $J^{(C)}=J^{(A)}$ and $ R^{(C)}=R^{(A)}$, i.e., two strongly-coupled qubits show the same behaviour as a single qubit. This can be intuitively understood looking at the spectrum of device \textbf{C}. As $g\to0$, the eigenstates  $\ket{\varepsilon_+}$ and $\ket{\varepsilon_-}$ become degenerate, and one is effectively left with qubit transitions of the same energy, $\varepsilon$. We must emphasize that while this limit makes mathematical sense, the global description of the two-qubit system breaks down for small $g$~\cite{Hofer2017} and one should consider again the model for device \textbf{B} for weakly interacting qubits.\par
Finally, we note that various two-qubit devices have been previously considered in \cite{Werlang2014,Miranda2017,Karg2019,Upadhyay2021}, and in Refs.~\cite{Miranda2017,Upadhyay2021} a trade-off between heat current and rectification was also noticed.  Furthermore,  in Ref.~\cite{Werlang2014,Miranda2017} the maximisation of $R$ was considered. Our considerations enable a more general optimisation within the $R-J$  plane, and  to operationally compare different rectifiers, as further discussed in the following section. \ref{sec:charac}.

\section{Comparing the performance of the three models}
\label{sec:charac}

\begin{figure}[t]
    \centering
    \includegraphics[width=0.95\columnwidth]{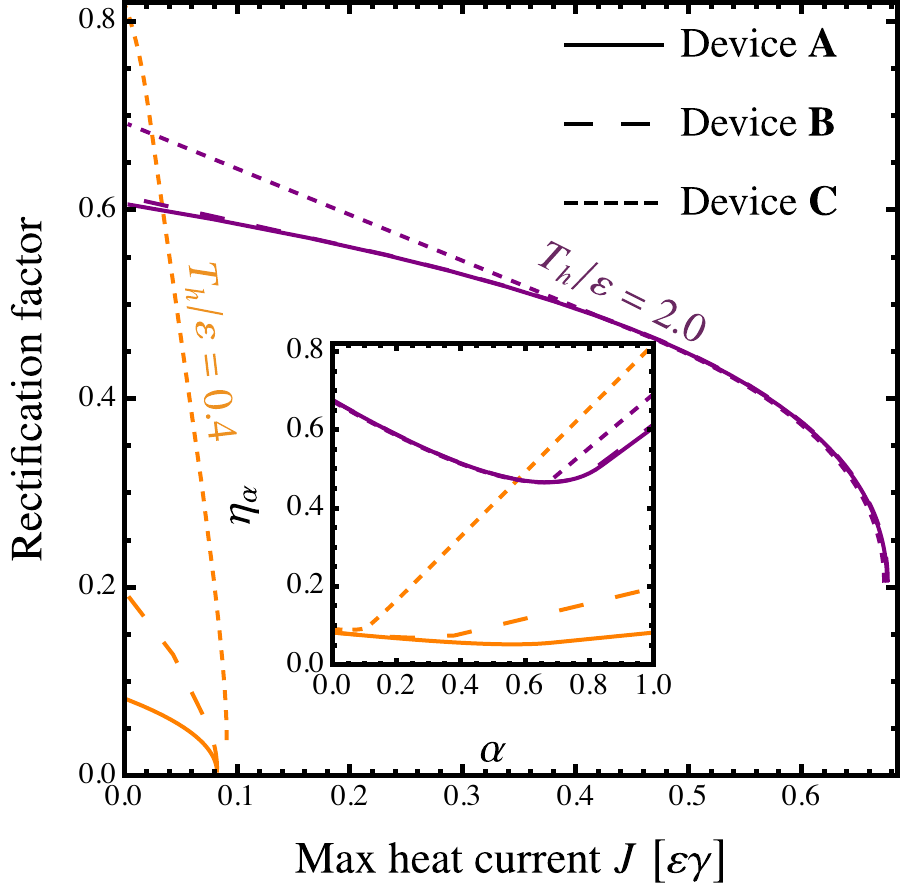}
    \caption{Rectification-Current Pareto-optimal fronts of each device by optimizing the COP $\eta_\alpha$ over all parameters' range relevant for each device; $\chi$ for device \textbf{A}, $\chi$, $\delta$ and $g$ for device \textbf{B} , $\chi$ and $g$ for device \textbf{C}, for two different hot temperatures, $T_h/\varepsilon = 2.0$ (purple) and $T_h/\varepsilon = 0.4$ (orange). The cold temperature is fixed to $T_c/\varepsilon = 0.01$, as well as $\gamma/\varepsilon=0.001$ is set constant. The inset shows optimal (maximum) value $\eta_\alpha$ as a function of $\alpha$, see Eq.~\eqref{eq:eta}. Optimization ranges: $0\leq \delta/\varepsilon\leq 0.1$, $0\leq \chi\leq1$, $0\leq g\leq0.05$ (weak coupling) and $0.1\leq g \leq 0.95$ (strong coupling).}
    \label{fig:pareto}
\end{figure}


\begin{figure*}
\centering
\begin{subfigure}
  \centering
  \includegraphics[width=.459\linewidth]{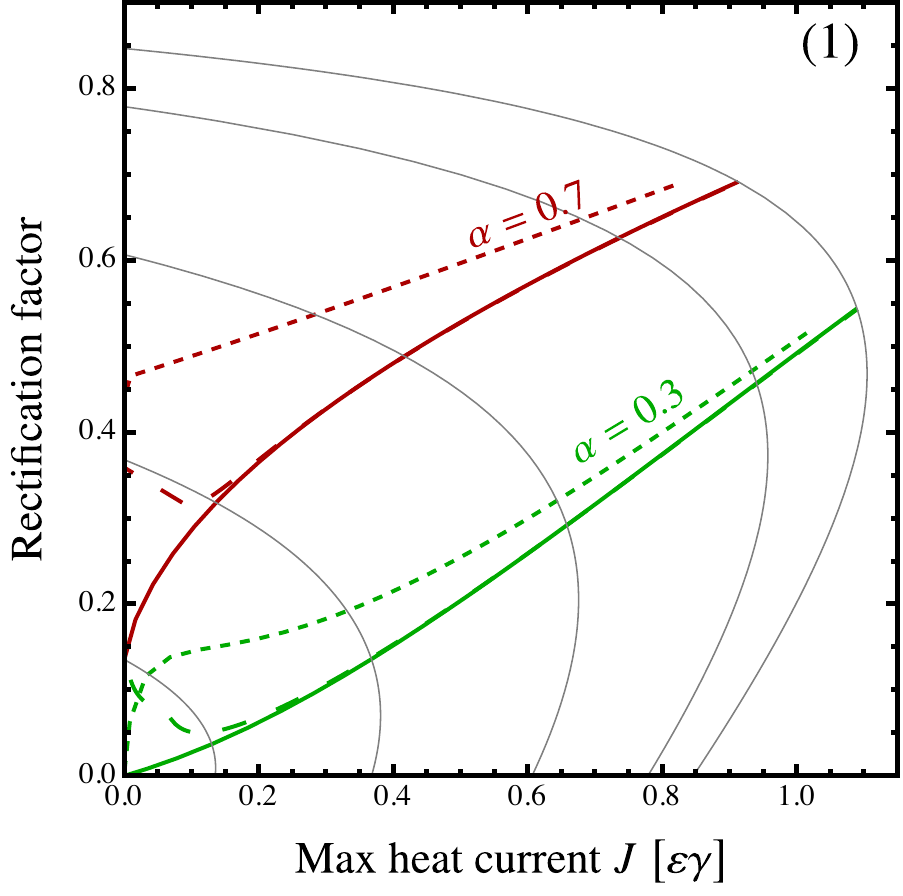}
\end{subfigure}%
\begin{subfigure}
  \centering
  \includegraphics[width=.46\linewidth]{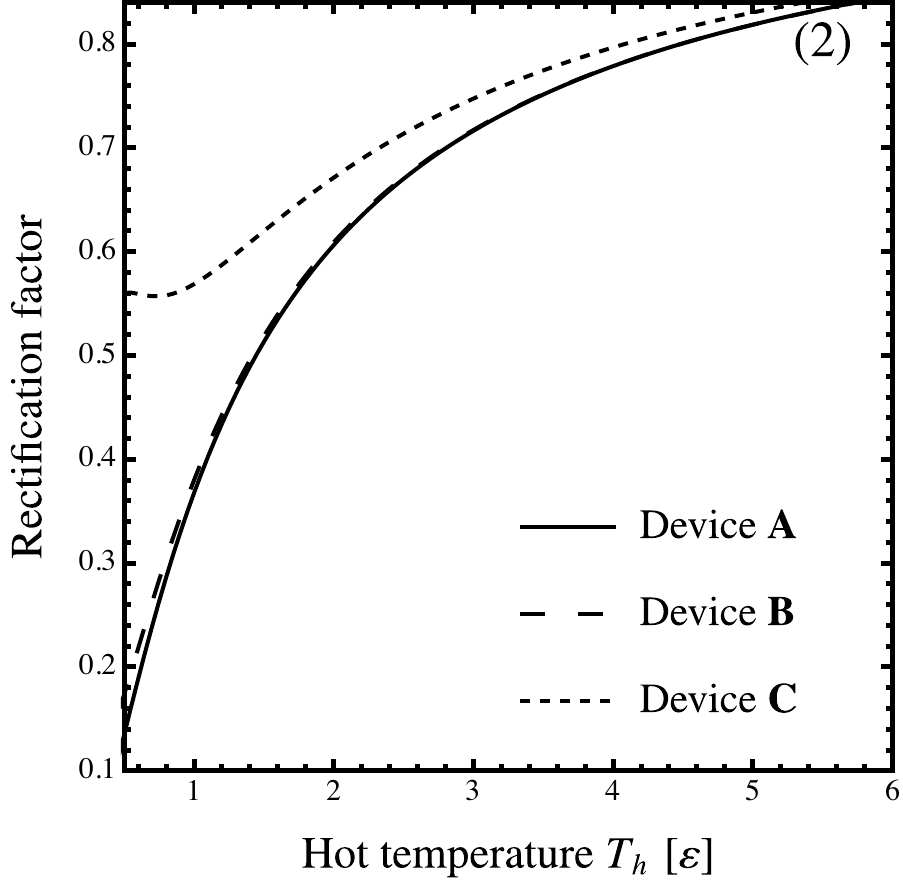}
\end{subfigure}
\caption{Performance of the three devices as a function of temperature gradient. Panel (1): Parametric plot of the rectification factor versus the maximum heat current as a function of asymmetry parameter $\chi$ for fixed $T_c/\varepsilon=0.01$ and $0.1\leq T_h/\varepsilon\leq 6$. Red and green sets of curves correspond to $\alpha = 0.7$ and 0.3 respectively. The solid curves correspond to device \textbf{A} (reproduced from Fig.~\ref{fig:sing} for clarity). The long-dashed and short-dashed curves correspond to devices \textbf{B} and \textbf{C}, respectively.  $\eta_\alpha$ is optimized over $\delta$, $g$ and $\chi$ for \textbf{B}, and $g$ and $\chi$ for \textbf{C}. Panel (2): maximum achievable rectification for the three devices as a function of hot temperature $T_h$ for fixed $T_c/\varepsilon = 0.01 $, given that $J\,[\gamma\varepsilon]\geq1$. At large $T_h$ (corresponding to a large temperature bias for a fixed $T_c$), advantage of \textbf{B} and \textbf{C} over device \textbf{A} decreases to 0. The single-qubit rectifier sets lower bounds on heat conduction and heat rectification. 
The other parameters are $\varepsilon=1$, $\gamma/\varepsilon=0.001$, $g/\varepsilon=0.8$ (strong coupling), $g/\varepsilon=0.1$ (weak coupling) and $T_c/\varepsilon=0.01$.}
\label{fig:optimal}
\end{figure*}

In the previous sections, we showed that any advantage one device holds over another in heat conduction or heat rectification strongly depends on the parameters with which they are operated. While this analysis allowed us to conclude that the performance of one device \textit{can} be better than that of another, it does not allow us to make a stronger statement about the overall performance of one device over another.

To address this question, we follow a procedure similar to the one for Fig. \ref{fig:sing}. For a fixed temperature,
we optimize $\eta_\alpha$ for all $0\leq\alpha\leq1$, over all the relevant parameters of the corresponding devices. We then plot the  rectification factor $R$ vs the maximum current $J$, obtaining Pareto fronts for each device, presented in Fig. \ref{fig:pareto}. Note that these fronts are Pareto-optimal, in the sense that rectification cannot be increased without sacrificing heat conduction, and vice-versa. The procedure is then repeated for a higher temperature, giving another set of three Pareto fronts.

We find that in both sets of Pareto fronts, the one corresponding to device \textbf{C} (dashed) lies above (or on top of) the other fronts, and the one corresponding to \textbf{B} (long-dashed) lies in the middle. This means that for any given heat current that is required, device \textbf{C} gives the highest rectification factors. Since, the optimization is performed for all $\alpha$, this result is independent of its value. This unambiguously demonstrates that device \textbf{C} is superior compared to devices \textbf{A} and \textbf{B} for heat conduction and heat rectification. The inset in Fig.~\ref{fig:pareto} shows the maximum value of $\eta_\alpha$ as a function of $\alpha$. For both choices of $T_h$, there is a specific value of $\alpha$ where there advantage of device \textbf{C} starts over devices \textbf{A} and \textbf{B}. Furthermore, the high temperature (purple) curves show that for increasing hot temperatures, there is a clear decrease in the advantage. We further investigate the performance of these three devices as a function of the temperature gradient in Fig.~\ref{fig:optimal}.

Similarly to Fig. \ref{fig:sing}, we show in Fig. \ref{fig:optimal} panel (1) the corresponding curves of optimal operation for devices \textbf{B} and \textbf{C} for increasing temperature gradients. With dashed and long-dashed curves, we show optimal lines of $\{R^{(B)}, J^{(B)}\}$ and $\{R^{(C)}, J^{(C)}\}$; at every point in each line, $\eta_\alpha$ is optimized for $\alpha = 0.3$ (green), and $\alpha = 0.7$ (red). The plot clearly demonstrates that devices \textbf{B} and \textbf{C} have a higher rectification factor when operated optimally, compared to device \textbf{A} at low temperatures. The advantage of device \textbf{C} over device \textbf{A} remains at much larger temperature gradients, but also tends to decrease, and to reach the single-qubit device which serves as a representative lower-bound. In panel (2), we further investigate this temperature behaviour. We show the maximal achievable rectification factor (optimizing over all relevant parameters) for the three devices as a function of the hot temperature, for a fix cold temperature and a given minimum heat current. We again find that device \textbf{C} holds an advantage, which decreases with increasing temperature and the curves merge at high temperatures.\par
Analytically, one can demonstrate that $R^{(B)}, R^{(C)} \rightarrow R^{(A)}$ for $T_h \rightarrow \infty$ as follows. In the limit of a high temperature bias, $n_B(E_k, T_h)\gg 1$ and $n_B(E_k, T_h) \gg n_B(E_k, T_c)$, the heat currents $J^{(B)}, J^{(C)}$ and rectification factors $R^{(B)}, R^{(C)}$ approach the limiting values $J^{(A)}_{hc/ch} \approx \pm 2\varepsilon \gamma (1\pm\chi)$ and $R^{(A)}\approx \lvert \chi\rvert$. This explains the convergence of the curves in Figs. \ref{fig:pareto} and \ref{fig:optimal}.

\section{Conclusion}

We presented a general method for assessing the performance of heat rectifiers which involves mapping the trade-off between the rectification factor $R$ and the maximal heat current $J$. To do so, we introduced a general coefficient of performance (COP), which can be optimized over the relevant parameters of the system allowing us to draw Pareto fronts and to meaningfully compare the performance of different rectifiers. We then illustrated the relevance of these ideas on three minimal models for nanoscale heat rectifiers and demonstrated that a strongly-coupled two-qubit device operates better than a single-qubit or a weakly-coupled two qubit device, both for heat rectification and heat conduction.
\par
Our proposed method and analysis can be readily applied to any physical device in order to characterize its performance as a heat rectifier. Hence, we believe that this approach to heat rectification might help identifying which type of systems are best suited to be operated as heat rectifiers. Our results on two-qubit devices suggest that strongly-interacting composite systems, may be beneficial for heat rectification. From there follows an open question about the \textit{optimal} device and its corresponding \textit{optimal} non-linear energy spectrum. This question was recently investigated in Refs.~\cite{Pereira2019, Pereira2019II} considering classical and quantum systems (Ising models) focusing on rectification alone. Our work opens the way to determine optimal device from a fundamental and practical points of view, by also taking into account their heat conduction properties. Another open question  concerns the possible advantage of a rectifier that would be operated in the quantum regime compared to a semi-classical regime. Indeed, the Pareto fronts in this work demonstrated the better performance of strongly-coupled qubits. However, this device can be modelled with a semi-classical model with rate equations describing the dynamics of the four energy eigenstates. Hence, the role of quantum coherence or quantum correlations for a better device in our approach remains an open question of particular importance for determining a possible quantum advantage. Our approach provides a solid framework to tackle these open questions in a practical, systematic and rigorous way.

\section*{Acknowledgements}

All authors acknowledge support from the Swiss National Science Foundation. SK and GH were supported by the starting grant PRIMA PR00P2\_179748, MPL by the Ambizione Grant No. PZ00P2-186067 and NB by NCCR SwissMAP.

\bibliography{References.bib}

\begin{thebibliography}{71}%
\makeatletter
\providecommand \@ifxundefined [1]{%
 \@ifx{#1\undefined}
}%
\providecommand \@ifnum [1]{%
 \ifnum #1\expandafter \@firstoftwo
 \else \expandafter \@secondoftwo
 \fi
}%
\providecommand \@ifx [1]{%
 \ifx #1\expandafter \@firstoftwo
 \else \expandafter \@secondoftwo
 \fi
}%
\providecommand \natexlab [1]{#1}%
\providecommand \enquote  [1]{``#1''}%
\providecommand \bibnamefont  [1]{#1}%
\providecommand \bibfnamefont [1]{#1}%
\providecommand \citenamefont [1]{#1}%
\providecommand \href@noop [0]{\@secondoftwo}%
\providecommand \href [0]{\begingroup \@sanitize@url \@href}%
\providecommand \@href[1]{\@@startlink{#1}\@@href}%
\providecommand \@@href[1]{\endgroup#1\@@endlink}%
\providecommand \@sanitize@url [0]{\catcode `\\12\catcode `\$12\catcode
  `\&12\catcode `\#12\catcode `\^12\catcode `\_12\catcode `\%12\relax}%
\providecommand \@@startlink[1]{}%
\providecommand \@@endlink[0]{}%
\providecommand \url  [0]{\begingroup\@sanitize@url \@url }%
\providecommand \@url [1]{\endgroup\@href {#1}{\urlprefix }}%
\providecommand \urlprefix  [0]{URL }%
\providecommand \Eprint [0]{\href }%
\providecommand \doibase [0]{http://dx.doi.org/}%
\providecommand \selectlanguage [0]{\@gobble}%
\providecommand \bibinfo  [0]{\@secondoftwo}%
\providecommand \bibfield  [0]{\@secondoftwo}%
\providecommand \translation [1]{[#1]}%
\providecommand \BibitemOpen [0]{}%
\providecommand \bibitemStop [0]{}%
\providecommand \bibitemNoStop [0]{.\EOS\space}%
\providecommand \EOS [0]{\spacefactor3000\relax}%
\providecommand \BibitemShut  [1]{\csname bibitem#1\endcsname}%
\let\auto@bib@innerbib\@empty
\bibitem [{\citenamefont {Giazotto}\ \emph {et~al.}(2006)\citenamefont
  {Giazotto}, \citenamefont {Heikkil\"{a}}, \citenamefont {Luukanen},
  \citenamefont {Savin},\ and\ \citenamefont {Pekola}}]{Giazotto2006}%
  \BibitemOpen
  \bibfield  {author} {\bibinfo {author} {\bibfnamefont {F.}~\bibnamefont
  {Giazotto}}, \bibinfo {author} {\bibfnamefont {T.~T.}\ \bibnamefont
  {Heikkil\"{a}}}, \bibinfo {author} {\bibfnamefont {A.}~\bibnamefont
  {Luukanen}}, \bibinfo {author} {\bibfnamefont {A.~M.}\ \bibnamefont {Savin}},
  \ and\ \bibinfo {author} {\bibfnamefont {J.~P.}\ \bibnamefont {Pekola}},\
  }\href {\doibase 10.1103/revmodphys.78.217} {\bibfield  {journal} {\bibinfo
  {journal} {Rev. Mod. Phys.}\ }\textbf {\bibinfo {volume} {78}},\ \bibinfo
  {pages} {217} (\bibinfo {year} {2006})}\BibitemShut {NoStop}%
\bibitem [{\citenamefont {Roberts}\ and\ \citenamefont
  {Walker}(2011)}]{Roberts2011}%
  \BibitemOpen
  \bibfield  {author} {\bibinfo {author} {\bibfnamefont {N.~A.}\ \bibnamefont
  {Roberts}}\ and\ \bibinfo {author} {\bibfnamefont {D.~G.}\ \bibnamefont
  {Walker}},\ }\href {\doibase 10.1016/j.ijthermalsci.2010.12.004} {\bibfield
  {journal} {\bibinfo  {journal} {Int. J. Therm. Sci.}\ }\textbf {\bibinfo
  {volume} {50}},\ \bibinfo {pages} {648} (\bibinfo {year} {2011})}\BibitemShut
  {NoStop}%
\bibitem [{\citenamefont {Li}\ \emph {et~al.}(2012)\citenamefont {Li},
  \citenamefont {Ren}, \citenamefont {Wang}, \citenamefont {Zhang},
  \citenamefont {H\"{a}nggi},\ and\ \citenamefont {Li}}]{Li2012}%
  \BibitemOpen
  \bibfield  {author} {\bibinfo {author} {\bibfnamefont {N.}~\bibnamefont
  {Li}}, \bibinfo {author} {\bibfnamefont {J.}~\bibnamefont {Ren}}, \bibinfo
  {author} {\bibfnamefont {L.}~\bibnamefont {Wang}}, \bibinfo {author}
  {\bibfnamefont {G.}~\bibnamefont {Zhang}}, \bibinfo {author} {\bibfnamefont
  {P.}~\bibnamefont {H\"{a}nggi}}, \ and\ \bibinfo {author} {\bibfnamefont
  {B.}~\bibnamefont {Li}},\ }\href
  {https://journals.aps.org/rmp/abstract/10.1103/RevModPhys.84.1045} {\bibfield
   {journal} {\bibinfo  {journal} {Rev. Mod. Phys.}\ }\textbf {\bibinfo
  {volume} {84}},\ \bibinfo {pages} {1045} (\bibinfo {year}
  {2012})}\BibitemShut {NoStop}%
\bibitem [{\citenamefont {Pekola}(2015)}]{Pekola2015}%
  \BibitemOpen
  \bibfield  {author} {\bibinfo {author} {\bibfnamefont {J.~P.}\ \bibnamefont
  {Pekola}},\ }\href {\doibase 10.1038/nphys3169} {\bibfield  {journal}
  {\bibinfo  {journal} {Nat. Phys.}\ }\textbf {\bibinfo {volume} {11}},\
  \bibinfo {pages} {118} (\bibinfo {year} {2015})}\BibitemShut {NoStop}%
\bibitem [{\citenamefont {Terraneo}\ \emph {et~al.}(2002)\citenamefont
  {Terraneo}, \citenamefont {Peyrard},\ and\ \citenamefont
  {Casati}}]{Terraneo2002}%
  \BibitemOpen
  \bibfield  {author} {\bibinfo {author} {\bibfnamefont {M.}~\bibnamefont
  {Terraneo}}, \bibinfo {author} {\bibfnamefont {M.}~\bibnamefont {Peyrard}}, \
  and\ \bibinfo {author} {\bibfnamefont {G.}~\bibnamefont {Casati}},\ }\href
  {\doibase 10.1103/PhysRevLett.88.094302} {\bibfield  {journal} {\bibinfo
  {journal} {Phys. Rev. Lett.}\ }\textbf {\bibinfo {volume} {88}},\ \bibinfo
  {pages} {094302} (\bibinfo {year} {2002})}\BibitemShut {NoStop}%
\bibitem [{\citenamefont {Li}\ \emph {et~al.}(2004)\citenamefont {Li},
  \citenamefont {Wang},\ and\ \citenamefont {Casati}}]{Li2004}%
  \BibitemOpen
  \bibfield  {author} {\bibinfo {author} {\bibfnamefont {B.}~\bibnamefont
  {Li}}, \bibinfo {author} {\bibfnamefont {L.}~\bibnamefont {Wang}}, \ and\
  \bibinfo {author} {\bibfnamefont {G.}~\bibnamefont {Casati}},\ }\href
  {\doibase 10.1103/PhysRevLett.93.184301} {\bibfield  {journal} {\bibinfo
  {journal} {Phys. Rev. Lett.}\ }\textbf {\bibinfo {volume} {93}},\ \bibinfo
  {pages} {184301} (\bibinfo {year} {2004})}\BibitemShut {NoStop}%
\bibitem [{\citenamefont {Li}\ \emph {et~al.}(2006)\citenamefont {Li},
  \citenamefont {Wang},\ and\ \citenamefont {Casati}}]{Li2006}%
  \BibitemOpen
  \bibfield  {author} {\bibinfo {author} {\bibfnamefont {B.}~\bibnamefont
  {Li}}, \bibinfo {author} {\bibfnamefont {L.}~\bibnamefont {Wang}}, \ and\
  \bibinfo {author} {\bibfnamefont {G.}~\bibnamefont {Casati}},\ }\href
  {https://doi.org/10.1063/1.2191730} {\bibfield  {journal} {\bibinfo
  {journal} {Appl Phys. Lett.}\ }\textbf {\bibinfo {volume} {88}},\ \bibinfo
  {pages} {143501} (\bibinfo {year} {2006})}\BibitemShut {NoStop}%
\bibitem [{\citenamefont {Segal}\ and\ \citenamefont
  {Nitzan}(2005)}]{Segal2005}%
  \BibitemOpen
  \bibfield  {author} {\bibinfo {author} {\bibfnamefont {D.}~\bibnamefont
  {Segal}}\ and\ \bibinfo {author} {\bibfnamefont {A.}~\bibnamefont {Nitzan}},\
  }\href {https://doi.org/10.1103/physrevlett.94.034301} {\bibfield  {journal}
  {\bibinfo  {journal} {Phys. Rev. Lett.}\ }\textbf {\bibinfo {volume} {94}}
  (\bibinfo {year} {2005})}\BibitemShut {NoStop}%
\bibitem [{\citenamefont {Wu}\ and\ \citenamefont {Segal}(2009)}]{Wu2009}%
  \BibitemOpen
  \bibfield  {author} {\bibinfo {author} {\bibfnamefont {L.-A.}\ \bibnamefont
  {Wu}}\ and\ \bibinfo {author} {\bibfnamefont {D.}~\bibnamefont {Segal}},\
  }\href {\doibase 10.1103/PhysRevLett.102.095503} {\bibfield  {journal}
  {\bibinfo  {journal} {Phys. Rev. Lett.}\ }\textbf {\bibinfo {volume} {102}},\
  \bibinfo {pages} {095503} (\bibinfo {year} {2009})}\BibitemShut {NoStop}%
\bibitem [{\citenamefont {Benenti}\ \emph {et~al.}(2017)\citenamefont
  {Benenti}, \citenamefont {Casati}, \citenamefont {Saito},\ and\ \citenamefont
  {Whitney}}]{Benenti2017}%
  \BibitemOpen
  \bibfield  {author} {\bibinfo {author} {\bibfnamefont {G.}~\bibnamefont
  {Benenti}}, \bibinfo {author} {\bibfnamefont {G.}~\bibnamefont {Casati}},
  \bibinfo {author} {\bibfnamefont {K.}~\bibnamefont {Saito}}, \ and\ \bibinfo
  {author} {\bibfnamefont {R.~S.}\ \bibnamefont {Whitney}},\ }\href {\doibase
  10.1016/j.physrep.2017.05.008} {\bibfield  {journal} {\bibinfo  {journal}
  {Phys. Rep.}\ }\textbf {\bibinfo {volume} {694}},\ \bibinfo {pages} {1}
  (\bibinfo {year} {2017})}\BibitemShut {NoStop}%
\bibitem [{\citenamefont {Landi}\ \emph {et~al.}(2021)\citenamefont {Landi},
  \citenamefont {Poletti},\ and\ \citenamefont {Schaller}}]{Landi2021}%
  \BibitemOpen
  \bibfield  {author} {\bibinfo {author} {\bibfnamefont {G.~T.}\ \bibnamefont
  {Landi}}, \bibinfo {author} {\bibfnamefont {D.}~\bibnamefont {Poletti}}, \
  and\ \bibinfo {author} {\bibfnamefont {G.}~\bibnamefont {Schaller}},\ }\href
  {https://arxiv.org/abs/2104.14350} {\bibfield  {journal} {\bibinfo  {journal}
  {arXiv:2104.14350 (2021)}\ } (\bibinfo {year} {2021})}\BibitemShut {NoStop}%
\bibitem [{\citenamefont {Wang}\ \emph {et~al.}(2017)\citenamefont {Wang},
  \citenamefont {Hu}, \citenamefont {Takahashi}, \citenamefont {Zhang},
  \citenamefont {Takamatsu},\ and\ \citenamefont {Chen}}]{Wang2017}%
  \BibitemOpen
  \bibfield  {author} {\bibinfo {author} {\bibfnamefont {H.}~\bibnamefont
  {Wang}}, \bibinfo {author} {\bibfnamefont {S.}~\bibnamefont {Hu}}, \bibinfo
  {author} {\bibfnamefont {K.}~\bibnamefont {Takahashi}}, \bibinfo {author}
  {\bibfnamefont {X.}~\bibnamefont {Zhang}}, \bibinfo {author} {\bibfnamefont
  {H.}~\bibnamefont {Takamatsu}}, \ and\ \bibinfo {author} {\bibfnamefont
  {J.}~\bibnamefont {Chen}},\ }\href {\doibase 10.1038/ncomms15843} {\bibfield
  {journal} {\bibinfo  {journal} {Nat. Commun.}\ }\textbf {\bibinfo {volume}
  {8}},\ \bibinfo {pages} {15843} (\bibinfo {year} {2017})}\BibitemShut
  {NoStop}%
\bibitem [{\citenamefont {Saira}\ \emph {et~al.}(2007)\citenamefont {Saira},
  \citenamefont {Meschke}, \citenamefont {Giazotto}, \citenamefont {Savin},
  \citenamefont {M\"ott\"onen},\ and\ \citenamefont {Pekola}}]{Saira2007}%
  \BibitemOpen
  \bibfield  {author} {\bibinfo {author} {\bibfnamefont {O.-P.}\ \bibnamefont
  {Saira}}, \bibinfo {author} {\bibfnamefont {M.}~\bibnamefont {Meschke}},
  \bibinfo {author} {\bibfnamefont {F.}~\bibnamefont {Giazotto}}, \bibinfo
  {author} {\bibfnamefont {A.~M.}\ \bibnamefont {Savin}}, \bibinfo {author}
  {\bibfnamefont {M.}~\bibnamefont {M\"ott\"onen}}, \ and\ \bibinfo {author}
  {\bibfnamefont {J.~P.}\ \bibnamefont {Pekola}},\ }\href {\doibase
  10.1103/PhysRevLett.99.027203} {\bibfield  {journal} {\bibinfo  {journal}
  {Phys. Rev. Lett.}\ }\textbf {\bibinfo {volume} {99}},\ \bibinfo {pages}
  {027203} (\bibinfo {year} {2007})}\BibitemShut {NoStop}%
\bibitem [{\citenamefont {Ronzani}\ \emph {et~al.}(2018)\citenamefont
  {Ronzani}, \citenamefont {Karimi}, \citenamefont {Senior}, \citenamefont
  {Chang}, \citenamefont {Peltonen}, \citenamefont {Chen},\ and\ \citenamefont
  {Pekola}}]{Ronzani2018}%
  \BibitemOpen
  \bibfield  {author} {\bibinfo {author} {\bibfnamefont {A.}~\bibnamefont
  {Ronzani}}, \bibinfo {author} {\bibfnamefont {B.}~\bibnamefont {Karimi}},
  \bibinfo {author} {\bibfnamefont {J.}~\bibnamefont {Senior}}, \bibinfo
  {author} {\bibfnamefont {Y.-C.}\ \bibnamefont {Chang}}, \bibinfo {author}
  {\bibfnamefont {J.~T.}\ \bibnamefont {Peltonen}}, \bibinfo {author}
  {\bibfnamefont {C.}~\bibnamefont {Chen}}, \ and\ \bibinfo {author}
  {\bibfnamefont {J.~P.}\ \bibnamefont {Pekola}},\ }\href
  {https://doi.org/10.1038/s41567-018-0199-4} {\bibfield  {journal} {\bibinfo
  {journal} {Nat. Phys.}\ }\textbf {\bibinfo {volume} {14}},\ \bibinfo {pages}
  {991} (\bibinfo {year} {2018})}\BibitemShut {NoStop}%
\bibitem [{\citenamefont {Senior}\ \emph {et~al.}(2020)\citenamefont {Senior},
  \citenamefont {Gubaydullin}, \citenamefont {Karimi}, \citenamefont
  {Peltonen}, \citenamefont {Ankerhold},\ and\ \citenamefont
  {Pekola}}]{Senior2020}%
  \BibitemOpen
  \bibfield  {author} {\bibinfo {author} {\bibfnamefont {J.}~\bibnamefont
  {Senior}}, \bibinfo {author} {\bibfnamefont {A.}~\bibnamefont {Gubaydullin}},
  \bibinfo {author} {\bibfnamefont {B.}~\bibnamefont {Karimi}}, \bibinfo
  {author} {\bibfnamefont {J.~T.}\ \bibnamefont {Peltonen}}, \bibinfo {author}
  {\bibfnamefont {J.}~\bibnamefont {Ankerhold}}, \ and\ \bibinfo {author}
  {\bibfnamefont {J.~P.}\ \bibnamefont {Pekola}},\ }\href
  {https://doi.org/10.1038/s42005-020-0307-5} {\bibfield  {journal} {\bibinfo
  {journal} {Comm. Phys.}\ }\textbf {\bibinfo {volume} {3}} (\bibinfo {year}
  {2020})}\BibitemShut {NoStop}%
\bibitem [{\citenamefont {Gubaydullin}\ \emph {et~al.}(2022)\citenamefont
  {Gubaydullin}, \citenamefont {Thomas}, \citenamefont {Golubev}, \citenamefont
  {Lvov}, \citenamefont {Peltonen},\ and\ \citenamefont
  {Pekola}}]{Gubaydullin2022}%
  \BibitemOpen
  \bibfield  {author} {\bibinfo {author} {\bibfnamefont {A.}~\bibnamefont
  {Gubaydullin}}, \bibinfo {author} {\bibfnamefont {G.}~\bibnamefont {Thomas}},
  \bibinfo {author} {\bibfnamefont {D.~S.}\ \bibnamefont {Golubev}}, \bibinfo
  {author} {\bibfnamefont {D.}~\bibnamefont {Lvov}}, \bibinfo {author}
  {\bibfnamefont {J.~T.}\ \bibnamefont {Peltonen}}, \ and\ \bibinfo {author}
  {\bibfnamefont {J.~P.}\ \bibnamefont {Pekola}},\ }\href
  {https://doi.org/10.1038/s41467-022-29078-x} {\bibfield  {journal} {\bibinfo
  {journal} {Nat. Commun.}\ }\textbf {\bibinfo {volume} {13}} (\bibinfo {year}
  {2022})}\BibitemShut {NoStop}%
\bibitem [{\citenamefont {Mart{\'\i}nez-P{\'e}rez}\ \emph
  {et~al.}(2015)\citenamefont {Mart{\'\i}nez-P{\'e}rez}, \citenamefont
  {Fornieri},\ and\ \citenamefont {Giazotto}}]{Martinez2015}%
  \BibitemOpen
  \bibfield  {author} {\bibinfo {author} {\bibfnamefont {M.~J.}\ \bibnamefont
  {Mart{\'\i}nez-P{\'e}rez}}, \bibinfo {author} {\bibfnamefont
  {A.}~\bibnamefont {Fornieri}}, \ and\ \bibinfo {author} {\bibfnamefont
  {F.}~\bibnamefont {Giazotto}},\ }\href {\doibase 10.1038/nnano.2015.11}
  {\bibfield  {journal} {\bibinfo  {journal} {Nat. Nanotechnol.}\ }\textbf
  {\bibinfo {volume} {10}},\ \bibinfo {pages} {303} (\bibinfo {year}
  {2015})}\BibitemShut {NoStop}%
\bibitem [{\citenamefont {Strambini}\ \emph {et~al.}(2022)\citenamefont
  {Strambini}, \citenamefont {Spies}, \citenamefont {Ligato}, \citenamefont
  {Ili{\'c}}, \citenamefont {Rouco}, \citenamefont {Gonz{\'a}lez-Orellana},
  \citenamefont {Ilyn}, \citenamefont {Rogero}, \citenamefont {Bergeret},
  \citenamefont {Moodera}, \citenamefont {Virtanen}, \citenamefont
  {Heikkil{\"a}},\ and\ \citenamefont {Giazotto}}]{Strambini2022}%
  \BibitemOpen
  \bibfield  {author} {\bibinfo {author} {\bibfnamefont {E.}~\bibnamefont
  {Strambini}}, \bibinfo {author} {\bibfnamefont {M.}~\bibnamefont {Spies}},
  \bibinfo {author} {\bibfnamefont {N.}~\bibnamefont {Ligato}}, \bibinfo
  {author} {\bibfnamefont {S.}~\bibnamefont {Ili{\'c}}}, \bibinfo {author}
  {\bibfnamefont {M.}~\bibnamefont {Rouco}}, \bibinfo {author} {\bibfnamefont
  {C.}~\bibnamefont {Gonz{\'a}lez-Orellana}}, \bibinfo {author} {\bibfnamefont
  {M.}~\bibnamefont {Ilyn}}, \bibinfo {author} {\bibfnamefont {C.}~\bibnamefont
  {Rogero}}, \bibinfo {author} {\bibfnamefont {F.~S.}\ \bibnamefont
  {Bergeret}}, \bibinfo {author} {\bibfnamefont {J.~S.}\ \bibnamefont
  {Moodera}}, \bibinfo {author} {\bibfnamefont {P.}~\bibnamefont {Virtanen}},
  \bibinfo {author} {\bibfnamefont {T.~T.}\ \bibnamefont {Heikkil{\"a}}}, \
  and\ \bibinfo {author} {\bibfnamefont {F.}~\bibnamefont {Giazotto}},\ }\href
  {\doibase 10.1038/s41467-022-29990-2} {\bibfield  {journal} {\bibinfo
  {journal} {Nat. Commun.}\ }\textbf {\bibinfo {volume} {13}},\ \bibinfo
  {pages} {2431} (\bibinfo {year} {2022})}\BibitemShut {NoStop}%
\bibitem [{\citenamefont {Wu}\ \emph {et~al.}(2009)\citenamefont {Wu},
  \citenamefont {Yu},\ and\ \citenamefont {Segal}}]{Wu2009b}%
  \BibitemOpen
  \bibfield  {author} {\bibinfo {author} {\bibfnamefont {L.-A.}\ \bibnamefont
  {Wu}}, \bibinfo {author} {\bibfnamefont {C.~X.}\ \bibnamefont {Yu}}, \ and\
  \bibinfo {author} {\bibfnamefont {D.}~\bibnamefont {Segal}},\ }\href
  {\doibase 10.1103/PhysRevE.80.041103} {\bibfield  {journal} {\bibinfo
  {journal} {Phys. Rev. E}\ }\textbf {\bibinfo {volume} {80}},\ \bibinfo
  {pages} {041103} (\bibinfo {year} {2009})}\BibitemShut {NoStop}%
\bibitem [{\citenamefont {Silva}\ \emph {et~al.}(2020)\citenamefont {Silva},
  \citenamefont {Landi}, \citenamefont {Drumond},\ and\ \citenamefont
  {Pereira}}]{Silva2020}%
  \BibitemOpen
  \bibfield  {author} {\bibinfo {author} {\bibfnamefont {S.~H.~S.}\
  \bibnamefont {Silva}}, \bibinfo {author} {\bibfnamefont {G.~T.}\ \bibnamefont
  {Landi}}, \bibinfo {author} {\bibfnamefont {R.~C.}\ \bibnamefont {Drumond}},
  \ and\ \bibinfo {author} {\bibfnamefont {E.}~\bibnamefont {Pereira}},\ }\href
  {\doibase 10.1103/PhysRevE.102.062146} {\bibfield  {journal} {\bibinfo
  {journal} {Phys. Rev. E}\ }\textbf {\bibinfo {volume} {102}},\ \bibinfo
  {pages} {062146} (\bibinfo {year} {2020})}\BibitemShut {NoStop}%
\bibitem [{\citenamefont {L\'opez}\ and\ \citenamefont
  {S\'anchez}(2013)}]{Lopez2013}%
  \BibitemOpen
  \bibfield  {author} {\bibinfo {author} {\bibfnamefont {R.}~\bibnamefont
  {L\'opez}}\ and\ \bibinfo {author} {\bibfnamefont {D.}~\bibnamefont
  {S\'anchez}},\ }\href {\doibase 10.1103/PhysRevB.88.045129} {\bibfield
  {journal} {\bibinfo  {journal} {Phys. Rev. B}\ }\textbf {\bibinfo {volume}
  {88}},\ \bibinfo {pages} {045129} (\bibinfo {year} {2013})}\BibitemShut
  {NoStop}%
\bibitem [{\citenamefont {Defaveri}\ and\ \citenamefont
  {Anteneodo}(2021)}]{Defaveri2021}%
  \BibitemOpen
  \bibfield  {author} {\bibinfo {author} {\bibfnamefont {L.}~\bibnamefont
  {Defaveri}}\ and\ \bibinfo {author} {\bibfnamefont {C.}~\bibnamefont
  {Anteneodo}},\ }\href {https://doi.org/10.1103/physreve.104.014106}
  {\bibfield  {journal} {\bibinfo  {journal} {Phys. Rev. E}\ }\textbf {\bibinfo
  {volume} {104}} (\bibinfo {year} {2021})}\BibitemShut {NoStop}%
\bibitem [{\citenamefont {Kalantar}\ \emph {et~al.}(2021)\citenamefont
  {Kalantar}, \citenamefont {Agarwalla},\ and\ \citenamefont
  {Segal}}]{Kalantar2021}%
  \BibitemOpen
  \bibfield  {author} {\bibinfo {author} {\bibfnamefont {N.}~\bibnamefont
  {Kalantar}}, \bibinfo {author} {\bibfnamefont {B.~K.}\ \bibnamefont
  {Agarwalla}}, \ and\ \bibinfo {author} {\bibfnamefont {D.}~\bibnamefont
  {Segal}},\ }\href {https://doi.org/10.1103/physreve.103.052130} {\bibfield
  {journal} {\bibinfo  {journal} {Phys. Rev. E}\ }\textbf {\bibinfo {volume}
  {103}} (\bibinfo {year} {2021})}\BibitemShut {NoStop}%
\bibitem [{\citenamefont {Man}\ \emph {et~al.}(2016)\citenamefont {Man},
  \citenamefont {An},\ and\ \citenamefont {Xia}}]{Man2016}%
  \BibitemOpen
  \bibfield  {author} {\bibinfo {author} {\bibfnamefont {Z.-X.}\ \bibnamefont
  {Man}}, \bibinfo {author} {\bibfnamefont {N.~B.}\ \bibnamefont {An}}, \ and\
  \bibinfo {author} {\bibfnamefont {Y.-J.}\ \bibnamefont {Xia}},\ }\href
  {\doibase 10.1103/PhysRevE.94.042135} {\bibfield  {journal} {\bibinfo
  {journal} {Phys. Rev. E}\ }\textbf {\bibinfo {volume} {94}},\ \bibinfo
  {pages} {042135} (\bibinfo {year} {2016})}\BibitemShut {NoStop}%
\bibitem [{\citenamefont {Mascarenhas}\ \emph {et~al.}(2016)\citenamefont
  {Mascarenhas}, \citenamefont {Santos}, \citenamefont {Auff{\`{e}}ves},\ and\
  \citenamefont {Gerace}}]{Mascarenhas2016}%
  \BibitemOpen
  \bibfield  {author} {\bibinfo {author} {\bibfnamefont {E.}~\bibnamefont
  {Mascarenhas}}, \bibinfo {author} {\bibfnamefont {M.~F.}\ \bibnamefont
  {Santos}}, \bibinfo {author} {\bibfnamefont {A.}~\bibnamefont
  {Auff{\`{e}}ves}}, \ and\ \bibinfo {author} {\bibfnamefont {D.}~\bibnamefont
  {Gerace}},\ }\href {https://doi.org/10.1103/physreva.93.043821} {\bibfield
  {journal} {\bibinfo  {journal} {Phys. Rev. A}\ }\textbf {\bibinfo {volume}
  {93}} (\bibinfo {year} {2016})}\BibitemShut {NoStop}%
\bibitem [{\citenamefont {Joulain}\ \emph {et~al.}(2016)\citenamefont
  {Joulain}, \citenamefont {Drevillon}, \citenamefont {Ezzahri},\ and\
  \citenamefont {Ordonez-Miranda}}]{Joulain2016}%
  \BibitemOpen
  \bibfield  {author} {\bibinfo {author} {\bibfnamefont {K.}~\bibnamefont
  {Joulain}}, \bibinfo {author} {\bibfnamefont {J.}~\bibnamefont {Drevillon}},
  \bibinfo {author} {\bibfnamefont {Y.}~\bibnamefont {Ezzahri}}, \ and\
  \bibinfo {author} {\bibfnamefont {J.}~\bibnamefont {Ordonez-Miranda}},\
  }\href {\doibase 10.1103/PhysRevLett.116.200601} {\bibfield  {journal}
  {\bibinfo  {journal} {Phys. Rev. Lett.}\ }\textbf {\bibinfo {volume} {116}},\
  \bibinfo {pages} {200601} (\bibinfo {year} {2016})}\BibitemShut {NoStop}%
\bibitem [{\citenamefont {Rossell{\'{o}}}\ \emph {et~al.}(2017)\citenamefont
  {Rossell{\'{o}}}, \citenamefont {L{\'{o}}pez},\ and\ \citenamefont
  {S{\'{a}}nchez}}]{Rossell2017}%
  \BibitemOpen
  \bibfield  {author} {\bibinfo {author} {\bibfnamefont {G.}~\bibnamefont
  {Rossell{\'{o}}}}, \bibinfo {author} {\bibfnamefont {R.}~\bibnamefont
  {L{\'{o}}pez}}, \ and\ \bibinfo {author} {\bibfnamefont {R.}~\bibnamefont
  {S{\'{a}}nchez}},\ }\href {https://doi.org/10.1103/physrevb.95.235404}
  {\bibfield  {journal} {\bibinfo  {journal} {Phys. Rev. B}\ }\textbf {\bibinfo
  {volume} {95}} (\bibinfo {year} {2017})}\BibitemShut {NoStop}%
\bibitem [{\citenamefont {Ordonez-Miranda}\ \emph {et~al.}(2017)\citenamefont
  {Ordonez-Miranda}, \citenamefont {Ezzahri},\ and\ \citenamefont
  {Joulain}}]{Miranda2017}%
  \BibitemOpen
  \bibfield  {author} {\bibinfo {author} {\bibfnamefont {J.}~\bibnamefont
  {Ordonez-Miranda}}, \bibinfo {author} {\bibfnamefont {Y.}~\bibnamefont
  {Ezzahri}}, \ and\ \bibinfo {author} {\bibfnamefont {K.}~\bibnamefont
  {Joulain}},\ }\href {https://doi.org/10.1103/PhysRevE.95.022128} {\bibfield
  {journal} {\bibinfo  {journal} {Phys. Rev. E}\ }\textbf {\bibinfo {volume}
  {95}},\ \bibinfo {pages} {022128} (\bibinfo {year} {2017})}\BibitemShut
  {NoStop}%
\bibitem [{\citenamefont {Tang}\ \emph {et~al.}(2018)\citenamefont {Tang},
  \citenamefont {Zhang},\ and\ \citenamefont {Wang}}]{Tang2018}%
  \BibitemOpen
  \bibfield  {author} {\bibinfo {author} {\bibfnamefont {G.}~\bibnamefont
  {Tang}}, \bibinfo {author} {\bibfnamefont {L.}~\bibnamefont {Zhang}}, \ and\
  \bibinfo {author} {\bibfnamefont {J.}~\bibnamefont {Wang}},\ }\href
  {https://link.aps.org/doi/10.1103/PhysRevB.97.224311} {\bibfield  {journal}
  {\bibinfo  {journal} {Phys. Rev. B}\ }\textbf {\bibinfo {volume} {97}},\
  \bibinfo {pages} {224311} (\bibinfo {year} {2018})}\BibitemShut {NoStop}%
\bibitem [{\citenamefont {Kargi}\ \emph {et~al.}(2019)\citenamefont {Kargi},
  \citenamefont {Naseem}, \citenamefont {Opatrn{\'{y}}}, \citenamefont
  {M{\"{u}}stecaplıoğlu},\ and\ \citenamefont {Kurizki}}]{Karg2019}%
  \BibitemOpen
  \bibfield  {author} {\bibinfo {author} {\bibfnamefont {C.}~\bibnamefont
  {Kargi}}, \bibinfo {author} {\bibfnamefont {M.~T.}\ \bibnamefont {Naseem}},
  \bibinfo {author} {\bibfnamefont {T.}~\bibnamefont {Opatrn{\'{y}}}}, \bibinfo
  {author} {\bibfnamefont {{\"{O}}.~E.}\ \bibnamefont
  {M{\"{u}}stecaplıoğlu}}, \ and\ \bibinfo {author} {\bibfnamefont
  {G.}~\bibnamefont {Kurizki}},\ }\href {\doibase 10.1103/PhysRevE.99.042121}
  {\bibfield  {journal} {\bibinfo  {journal} {Phys. Rev. E}\ }\textbf {\bibinfo
  {volume} {99}},\ \bibinfo {pages} {042121} (\bibinfo {year}
  {2019})}\BibitemShut {NoStop}%
\bibitem [{\citenamefont {Aligia}\ \emph {et~al.}(2020)\citenamefont {Aligia},
  \citenamefont {Daroca}, \citenamefont {Arrachea},\ and\ \citenamefont
  {Roura-Bas}}]{Aligia2020}%
  \BibitemOpen
  \bibfield  {author} {\bibinfo {author} {\bibfnamefont {A.~A.}\ \bibnamefont
  {Aligia}}, \bibinfo {author} {\bibfnamefont {D.~P.}\ \bibnamefont {Daroca}},
  \bibinfo {author} {\bibfnamefont {L.}~\bibnamefont {Arrachea}}, \ and\
  \bibinfo {author} {\bibfnamefont {P.}~\bibnamefont {Roura-Bas}},\ }\href
  {\doibase 10.1103/PhysRevB.101.075417} {\bibfield  {journal} {\bibinfo
  {journal} {Phys. Rev. B}\ }\textbf {\bibinfo {volume} {101}},\ \bibinfo
  {pages} {075417} (\bibinfo {year} {2020})}\BibitemShut {NoStop}%
\bibitem [{\citenamefont {Tesser}\ \emph {et~al.}(2022)\citenamefont {Tesser},
  \citenamefont {Bhandari}, \citenamefont {Erdman}, \citenamefont {Paladino},
  \citenamefont {Fazio},\ and\ \citenamefont {Taddei}}]{Tesser2021}%
  \BibitemOpen
  \bibfield  {author} {\bibinfo {author} {\bibfnamefont {L.}~\bibnamefont
  {Tesser}}, \bibinfo {author} {\bibfnamefont {B.}~\bibnamefont {Bhandari}},
  \bibinfo {author} {\bibfnamefont {P.~A.}\ \bibnamefont {Erdman}}, \bibinfo
  {author} {\bibfnamefont {E.}~\bibnamefont {Paladino}}, \bibinfo {author}
  {\bibfnamefont {R.}~\bibnamefont {Fazio}}, \ and\ \bibinfo {author}
  {\bibfnamefont {F.}~\bibnamefont {Taddei}},\ }\href
  {https://doi.org/10.1088/1367-2630/ac53b8} {\bibfield  {journal} {\bibinfo
  {journal} {New J. Phys.}\ }\textbf {\bibinfo {volume} {24}},\ \bibinfo
  {pages} {035001} (\bibinfo {year} {2022})}\BibitemShut {NoStop}%
\bibitem [{\citenamefont {Werlang}\ \emph {et~al.}(2014)\citenamefont
  {Werlang}, \citenamefont {Marchiori}, \citenamefont {Cornelio},\ and\
  \citenamefont {Valente}}]{Werlang2014}%
  \BibitemOpen
  \bibfield  {author} {\bibinfo {author} {\bibfnamefont {T.}~\bibnamefont
  {Werlang}}, \bibinfo {author} {\bibfnamefont {M.~A.}\ \bibnamefont
  {Marchiori}}, \bibinfo {author} {\bibfnamefont {M.~F.}\ \bibnamefont
  {Cornelio}}, \ and\ \bibinfo {author} {\bibfnamefont {D.}~\bibnamefont
  {Valente}},\ }\href {\doibase 10.1103/PhysRevE.89.062109} {\bibfield
  {journal} {\bibinfo  {journal} {Phys. Rev. E}\ }\textbf {\bibinfo {volume}
  {89}},\ \bibinfo {pages} {062109} (\bibinfo {year} {2014})}\BibitemShut
  {NoStop}%
\bibitem [{\citenamefont {Schuab}\ \emph {et~al.}(2016)\citenamefont {Schuab},
  \citenamefont {Pereira},\ and\ \citenamefont {Landi}}]{Schuab2016}%
  \BibitemOpen
  \bibfield  {author} {\bibinfo {author} {\bibfnamefont {L.}~\bibnamefont
  {Schuab}}, \bibinfo {author} {\bibfnamefont {E.}~\bibnamefont {Pereira}}, \
  and\ \bibinfo {author} {\bibfnamefont {G.~T.}\ \bibnamefont {Landi}},\ }\href
  {\doibase 10.1103/PhysRevE.94.042122} {\bibfield  {journal} {\bibinfo
  {journal} {Phys. Rev. E}\ }\textbf {\bibinfo {volume} {94}},\ \bibinfo
  {pages} {042122} (\bibinfo {year} {2016})}\BibitemShut {NoStop}%
\bibitem [{\citenamefont {S{\'{a}}nchez}\ \emph {et~al.}(2017)\citenamefont
  {S{\'{a}}nchez}, \citenamefont {Thierschmann},\ and\ \citenamefont
  {Molenkamp}}]{Sanchez2017}%
  \BibitemOpen
  \bibfield  {author} {\bibinfo {author} {\bibfnamefont {R.}~\bibnamefont
  {S{\'{a}}nchez}}, \bibinfo {author} {\bibfnamefont {H.}~\bibnamefont
  {Thierschmann}}, \ and\ \bibinfo {author} {\bibfnamefont {L.~W.}\
  \bibnamefont {Molenkamp}},\ }\href {\doibase 10.1088/1367-2630/aa8b94}
  {\bibfield  {journal} {\bibinfo  {journal} {New J. Phys.}\ }\textbf {\bibinfo
  {volume} {19}},\ \bibinfo {pages} {113040} (\bibinfo {year}
  {2017})}\BibitemShut {NoStop}%
\bibitem [{\citenamefont {Balachandran}\ \emph
  {et~al.}(2019{\natexlab{a}})\citenamefont {Balachandran}, \citenamefont
  {Clark}, \citenamefont {Goold},\ and\ \citenamefont
  {Poletti}}]{Balachandran2019}%
  \BibitemOpen
  \bibfield  {author} {\bibinfo {author} {\bibfnamefont {V.}~\bibnamefont
  {Balachandran}}, \bibinfo {author} {\bibfnamefont {S.~R.}\ \bibnamefont
  {Clark}}, \bibinfo {author} {\bibfnamefont {J.}~\bibnamefont {Goold}}, \ and\
  \bibinfo {author} {\bibfnamefont {D.}~\bibnamefont {Poletti}},\ }\href
  {\doibase 10.1103/PhysRevLett.123.020603} {\bibfield  {journal} {\bibinfo
  {journal} {Phys. Rev. Lett.}\ }\textbf {\bibinfo {volume} {123}},\ \bibinfo
  {pages} {020603} (\bibinfo {year} {2019}{\natexlab{a}})}\BibitemShut
  {NoStop}%
\bibitem [{\citenamefont {Balachandran}\ \emph
  {et~al.}(2019{\natexlab{b}})\citenamefont {Balachandran}, \citenamefont
  {Benenti}, \citenamefont {Pereira}, \citenamefont {Casati},\ and\
  \citenamefont {Poletti}}]{Balachandran2019II}%
  \BibitemOpen
  \bibfield  {author} {\bibinfo {author} {\bibfnamefont {V.}~\bibnamefont
  {Balachandran}}, \bibinfo {author} {\bibfnamefont {G.}~\bibnamefont
  {Benenti}}, \bibinfo {author} {\bibfnamefont {E.}~\bibnamefont {Pereira}},
  \bibinfo {author} {\bibfnamefont {G.}~\bibnamefont {Casati}}, \ and\ \bibinfo
  {author} {\bibfnamefont {D.}~\bibnamefont {Poletti}},\ }\href {\doibase
  10.1103/PhysRevE.99.032136} {\bibfield  {journal} {\bibinfo  {journal} {Phys.
  Rev. E}\ }\textbf {\bibinfo {volume} {99}},\ \bibinfo {pages} {032136}
  (\bibinfo {year} {2019}{\natexlab{b}})}\BibitemShut {NoStop}%
\bibitem [{\citenamefont {Iorio}\ \emph {et~al.}(2021)\citenamefont {Iorio},
  \citenamefont {Strambini}, \citenamefont {Haack}, \citenamefont {Campisi},\
  and\ \citenamefont {Giazotto}}]{Iorio2021}%
  \BibitemOpen
  \bibfield  {author} {\bibinfo {author} {\bibfnamefont {A.}~\bibnamefont
  {Iorio}}, \bibinfo {author} {\bibfnamefont {E.}~\bibnamefont {Strambini}},
  \bibinfo {author} {\bibfnamefont {G.}~\bibnamefont {Haack}}, \bibinfo
  {author} {\bibfnamefont {M.}~\bibnamefont {Campisi}}, \ and\ \bibinfo
  {author} {\bibfnamefont {F.}~\bibnamefont {Giazotto}},\ }\href
  {https://doi.org/10.1103/physrevapplied.15.054050} {\bibfield  {journal}
  {\bibinfo  {journal} {Phys. Rev. Appl.}\ }\textbf {\bibinfo {volume} {15}}
  (\bibinfo {year} {2021})}\BibitemShut {NoStop}%
\bibitem [{\citenamefont {Zhang}\ and\ \citenamefont {Su}(2021)}]{Zhang2021}%
  \BibitemOpen
  \bibfield  {author} {\bibinfo {author} {\bibfnamefont {Y.}~\bibnamefont
  {Zhang}}\ and\ \bibinfo {author} {\bibfnamefont {S.}~\bibnamefont {Su}},\
  }\href {\doibase 10.1016/j.physa.2021.126347} {\bibfield  {journal} {\bibinfo
   {journal} {Phys. A: Stat. Mech. Appl.}\ }\textbf {\bibinfo {volume} {584}},\
  \bibinfo {pages} {126347} (\bibinfo {year} {2021})}\BibitemShut {NoStop}%
\bibitem [{\citenamefont {Upadhyay}\ \emph {et~al.}(2021)\citenamefont
  {Upadhyay}, \citenamefont {Naseem}, \citenamefont {Marathe},\ and\
  \citenamefont {M\"ustecapl\ifmmode \imath \else \i
  \fi{}o\ifmmode~\breve{g}\else \u{g}\fi{}lu}}]{Upadhyay2021}%
  \BibitemOpen
  \bibfield  {author} {\bibinfo {author} {\bibfnamefont {V.}~\bibnamefont
  {Upadhyay}}, \bibinfo {author} {\bibfnamefont {M.~T.}\ \bibnamefont
  {Naseem}}, \bibinfo {author} {\bibfnamefont {R.}~\bibnamefont {Marathe}}, \
  and\ \bibinfo {author} {\bibfnamefont {O.~E.}\ \bibnamefont
  {M\"ustecapl\ifmmode \imath \else \i \fi{}o\ifmmode~\breve{g}\else
  \u{g}\fi{}lu}},\ }\href {\doibase 10.1103/PhysRevE.104.054137} {\bibfield
  {journal} {\bibinfo  {journal} {Phys. Rev. E}\ }\textbf {\bibinfo {volume}
  {104}},\ \bibinfo {pages} {054137} (\bibinfo {year} {2021})}\BibitemShut
  {NoStop}%
\bibitem [{\citenamefont {Guimar\~aes}\ \emph {et~al.}(2015)\citenamefont
  {Guimar\~aes}, \citenamefont {Landi},\ and\ \citenamefont
  {de~Oliveira}}]{Guimaraes2015}%
  \BibitemOpen
  \bibfield  {author} {\bibinfo {author} {\bibfnamefont {P.~H.}\ \bibnamefont
  {Guimar\~aes}}, \bibinfo {author} {\bibfnamefont {G.~T.}\ \bibnamefont
  {Landi}}, \ and\ \bibinfo {author} {\bibfnamefont {M.~J.}\ \bibnamefont
  {de~Oliveira}},\ }\href {\doibase 10.1103/PhysRevE.92.062120} {\bibfield
  {journal} {\bibinfo  {journal} {Phys. Rev. E}\ }\textbf {\bibinfo {volume}
  {92}},\ \bibinfo {pages} {062120} (\bibinfo {year} {2015})}\BibitemShut
  {NoStop}%
\bibitem [{\citenamefont {Sim{\'{o}}n}\ \emph {et~al.}(2021)\citenamefont
  {Sim{\'{o}}n}, \citenamefont {Ala{\~{n}}a}, \citenamefont {Pons},
  \citenamefont {Ruiz-Garc{\'{\i}}a},\ and\ \citenamefont {Muga}}]{Simn2021}%
  \BibitemOpen
  \bibfield  {author} {\bibinfo {author} {\bibfnamefont {M.~A.}\ \bibnamefont
  {Sim{\'{o}}n}}, \bibinfo {author} {\bibfnamefont {A.}~\bibnamefont
  {Ala{\~{n}}a}}, \bibinfo {author} {\bibfnamefont {M.}~\bibnamefont {Pons}},
  \bibinfo {author} {\bibfnamefont {A.}~\bibnamefont {Ruiz-Garc{\'{\i}}a}}, \
  and\ \bibinfo {author} {\bibfnamefont {J.~G.}\ \bibnamefont {Muga}},\ }\href
  {https://doi.org/10.1103/physreve.103.012134} {\bibfield  {journal} {\bibinfo
   {journal} {Phys. Rev. E}\ }\textbf {\bibinfo {volume} {103}} (\bibinfo
  {year} {2021})}\BibitemShut {NoStop}%
\bibitem [{\citenamefont {Romero-Bastida}\ and\ \citenamefont
  {Lindero-Hern{\'{a}}ndez}(2021)}]{RomeroBastida2021}%
  \BibitemOpen
  \bibfield  {author} {\bibinfo {author} {\bibfnamefont {M.}~\bibnamefont
  {Romero-Bastida}}\ and\ \bibinfo {author} {\bibfnamefont {M.}~\bibnamefont
  {Lindero-Hern{\'{a}}ndez}},\ }\href
  {https://doi.org/10.1103/physreve.104.044135} {\bibfield  {journal} {\bibinfo
   {journal} {Phys. Rev. E}\ }\textbf {\bibinfo {volume} {104}} (\bibinfo
  {year} {2021})}\BibitemShut {NoStop}%
\bibitem [{\citenamefont {Marcos-Vicioso}\ \emph
  {et~al.}(2018{\natexlab{a}})\citenamefont {Marcos-Vicioso}, \citenamefont
  {L{\'{o}}pez-Jurado}, \citenamefont {Ruiz-Garcia},\ and\ \citenamefont
  {S{\'{a}}nchez}}]{MarcosVicioso2018}%
  \BibitemOpen
  \bibfield  {author} {\bibinfo {author} {\bibfnamefont {A.}~\bibnamefont
  {Marcos-Vicioso}}, \bibinfo {author} {\bibfnamefont {C.}~\bibnamefont
  {L{\'{o}}pez-Jurado}}, \bibinfo {author} {\bibfnamefont {M.}~\bibnamefont
  {Ruiz-Garcia}}, \ and\ \bibinfo {author} {\bibfnamefont {R.}~\bibnamefont
  {S{\'{a}}nchez}},\ }\href {https://doi.org/10.1103/physrevb.98.035414}
  {\bibfield  {journal} {\bibinfo  {journal} {Phys. Rev. B}\ }\textbf {\bibinfo
  {volume} {98}} (\bibinfo {year} {2018}{\natexlab{a}})}\BibitemShut {NoStop}%
\bibitem [{\citenamefont {Goury}\ and\ \citenamefont
  {S{\'{a}}nchez}(2019)}]{Goury2019}%
  \BibitemOpen
  \bibfield  {author} {\bibinfo {author} {\bibfnamefont {D.}~\bibnamefont
  {Goury}}\ and\ \bibinfo {author} {\bibfnamefont {R.}~\bibnamefont
  {S{\'{a}}nchez}},\ }\href {\doibase 10.1063/1.5109100} {\bibfield  {journal}
  {\bibinfo  {journal} {Appl. Phys. Lett.}\ }\textbf {\bibinfo {volume}
  {115}},\ \bibinfo {pages} {092601} (\bibinfo {year} {2019})}\BibitemShut
  {NoStop}%
\bibitem [{\citenamefont {Poulsen}\ and\ \citenamefont
  {Zinner}()}]{Poulsen2022}%
  \BibitemOpen
  \bibfield  {author} {\bibinfo {author} {\bibfnamefont {K.}~\bibnamefont
  {Poulsen}}\ and\ \bibinfo {author} {\bibfnamefont {N.~T.}\ \bibnamefont
  {Zinner}},\ }\href {https://arxiv.org/abs/2203.12623} {\bibinfo  {journal}
  {arXiv:2203.12623 (2022)}\ }\BibitemShut {NoStop}%
\bibitem [{\citenamefont {Palafox}\ \emph {et~al.}()\citenamefont {Palafox},
  \citenamefont {Román-Ancheyta}, \citenamefont {Çakmak},\ and\ \citenamefont
  {M\"{u}stecaplıoğlu}}]{Palafox2022}%
  \BibitemOpen
\bibfield  {journal} {  }\bibfield  {author} {\bibinfo {author} {\bibfnamefont
  {S.}~\bibnamefont {Palafox}}, \bibinfo {author} {\bibfnamefont
  {R.}~\bibnamefont {Román-Ancheyta}}, \bibinfo {author} {\bibfnamefont
  {B.}~\bibnamefont {Çakmak}}, \ and\ \bibinfo {author} {\bibfnamefont
  {O.~E.}\ \bibnamefont {M\"{u}stecaplıoğlu}},\ }\href
  {https://arxiv.org/abs/2204.07060} {\bibinfo  {journal} {arXiv:2204.07060
  (2022)}\ }\BibitemShut {NoStop}%
\bibitem [{\citenamefont {Ruokola}\ \emph {et~al.}(2009)\citenamefont
  {Ruokola}, \citenamefont {Ojanen},\ and\ \citenamefont
  {Jauho}}]{Ruokola2009}%
  \BibitemOpen
\bibfield  {journal} {  }\bibfield  {author} {\bibinfo {author} {\bibfnamefont
  {T.}~\bibnamefont {Ruokola}}, \bibinfo {author} {\bibfnamefont
  {T.}~\bibnamefont {Ojanen}}, \ and\ \bibinfo {author} {\bibfnamefont {A.-P.}\
  \bibnamefont {Jauho}},\ }\href {\doibase 10.1103/PhysRevB.79.144306}
  {\bibfield  {journal} {\bibinfo  {journal} {Phys. Rev. B}\ }\textbf {\bibinfo
  {volume} {79}},\ \bibinfo {pages} {144306} (\bibinfo {year}
  {2009})}\BibitemShut {NoStop}%
\bibitem [{\citenamefont {Ruokola}\ and\ \citenamefont
  {Ojanen}(2011)}]{Ruokola2011}%
  \BibitemOpen
  \bibfield  {author} {\bibinfo {author} {\bibfnamefont {T.}~\bibnamefont
  {Ruokola}}\ and\ \bibinfo {author} {\bibfnamefont {T.}~\bibnamefont
  {Ojanen}},\ }\href {https://doi.org/10.1103/PhysRevB.83.241404} {\bibfield
  {journal} {\bibinfo  {journal} {Phys. Rev. B}\ }\textbf {\bibinfo {volume}
  {83}},\ \bibinfo {pages} {241404} (\bibinfo {year} {2011})}\BibitemShut
  {NoStop}%
\bibitem [{\citenamefont {Marcos-Vicioso}\ \emph
  {et~al.}(2018{\natexlab{b}})\citenamefont {Marcos-Vicioso}, \citenamefont
  {L{\'{o}}pez-Jurado}, \citenamefont {Ruiz-Garcia},\ and\ \citenamefont
  {S{\'{a}}nchez}}]{Marcos-Vicioso2018}%
  \BibitemOpen
  \bibfield  {author} {\bibinfo {author} {\bibfnamefont {A.}~\bibnamefont
  {Marcos-Vicioso}}, \bibinfo {author} {\bibfnamefont {C.}~\bibnamefont
  {L{\'{o}}pez-Jurado}}, \bibinfo {author} {\bibfnamefont {M.}~\bibnamefont
  {Ruiz-Garcia}}, \ and\ \bibinfo {author} {\bibfnamefont {R.}~\bibnamefont
  {S{\'{a}}nchez}},\ }\href {https://doi.org/10.1103/physrevb.98.035414}
  {\bibfield  {journal} {\bibinfo  {journal} {Phys. Rev. B}\ }\textbf {\bibinfo
  {volume} {98}} (\bibinfo {year} {2018}{\natexlab{b}})}\BibitemShut {NoStop}%
\bibitem [{\citenamefont {Riera-Campeny}\ \emph {et~al.}(2019)\citenamefont
  {Riera-Campeny}, \citenamefont {Mehboudi}, \citenamefont {Pons},\ and\
  \citenamefont {Sanpera}}]{Riera-Campeny2019}%
  \BibitemOpen
  \bibfield  {author} {\bibinfo {author} {\bibfnamefont {A.}~\bibnamefont
  {Riera-Campeny}}, \bibinfo {author} {\bibfnamefont {M.}~\bibnamefont
  {Mehboudi}}, \bibinfo {author} {\bibfnamefont {M.}~\bibnamefont {Pons}}, \
  and\ \bibinfo {author} {\bibfnamefont {A.}~\bibnamefont {Sanpera}},\ }\href
  {https://doi.org/10.1103/physreve.99.032126} {\bibfield  {journal} {\bibinfo
  {journal} {Phys. Rev. E}\ }\textbf {\bibinfo {volume} {99}} (\bibinfo {year}
  {2019})}\BibitemShut {NoStop}%
\bibitem [{\citenamefont {Bhandari}\ \emph {et~al.}(2021)\citenamefont
  {Bhandari}, \citenamefont {Erdman}, \citenamefont {Fazio}, \citenamefont
  {Paladino},\ and\ \citenamefont {Taddei}}]{Bhandari2021}%
  \BibitemOpen
  \bibfield  {author} {\bibinfo {author} {\bibfnamefont {B.}~\bibnamefont
  {Bhandari}}, \bibinfo {author} {\bibfnamefont {P.~A.}\ \bibnamefont
  {Erdman}}, \bibinfo {author} {\bibfnamefont {R.}~\bibnamefont {Fazio}},
  \bibinfo {author} {\bibfnamefont {E.}~\bibnamefont {Paladino}}, \ and\
  \bibinfo {author} {\bibfnamefont {F.}~\bibnamefont {Taddei}},\ }\href
  {https://doi.org/10.1103/physrevb.103.155434} {\bibfield  {journal} {\bibinfo
   {journal} {Phys. Rev. B}\ }\textbf {\bibinfo {volume} {103}} (\bibinfo
  {year} {2021})}\BibitemShut {NoStop}%
\bibitem [{\citenamefont {D{\'{\i}}az}\ and\ \citenamefont
  {S{\'{a}}nchez}(2021)}]{Daz2021}%
  \BibitemOpen
  \bibfield  {author} {\bibinfo {author} {\bibfnamefont {I.}~\bibnamefont
  {D{\'{\i}}az}}\ and\ \bibinfo {author} {\bibfnamefont {R.}~\bibnamefont
  {S{\'{a}}nchez}},\ }\href {\doibase 10.1088/1367-2630/ac4211} {\bibfield
  {journal} {\bibinfo  {journal} {New J. Phys.}\ }\textbf {\bibinfo {volume}
  {23}},\ \bibinfo {pages} {125006} (\bibinfo {year} {2021})}\BibitemShut
  {NoStop}%
\bibitem [{\citenamefont {Fornieri}\ \emph {et~al.}(2014)\citenamefont
  {Fornieri}, \citenamefont {Mart{\'{i}}nez-P{\'{e}}rez},\ and\ \citenamefont
  {Giazotto}}]{Fornieri2014}%
  \BibitemOpen
  \bibfield  {author} {\bibinfo {author} {\bibfnamefont {A.}~\bibnamefont
  {Fornieri}}, \bibinfo {author} {\bibfnamefont {M.~J.}\ \bibnamefont
  {Mart{\'{i}}nez-P{\'{e}}rez}}, \ and\ \bibinfo {author} {\bibfnamefont
  {F.}~\bibnamefont {Giazotto}},\ }\href
  {https://aip.scitation.org/doi/abs/10.1063/1.4875917} {\bibfield  {journal}
  {\bibinfo  {journal} {Appl. Phys. Lett.}\ }\textbf {\bibinfo {volume}
  {104}},\ \bibinfo {pages} {183108} (\bibinfo {year} {2014})}\BibitemShut
  {NoStop}%
\bibitem [{\citenamefont {S{\'{a}}nchez}\ \emph {et~al.}(2015)\citenamefont
  {S{\'{a}}nchez}, \citenamefont {Sothmann},\ and\ \citenamefont
  {Jordan}}]{Sanchez2015}%
  \BibitemOpen
  \bibfield  {author} {\bibinfo {author} {\bibfnamefont {R.}~\bibnamefont
  {S{\'{a}}nchez}}, \bibinfo {author} {\bibfnamefont {B.}~\bibnamefont
  {Sothmann}}, \ and\ \bibinfo {author} {\bibfnamefont {A.~N.}\ \bibnamefont
  {Jordan}},\ }\href {\doibase 10.1088/1367-2630/17/7/075006} {\bibfield
  {journal} {\bibinfo  {journal} {New J. Phys.}\ }\textbf {\bibinfo {volume}
  {17}},\ \bibinfo {pages} {075006} (\bibinfo {year} {2015})}\BibitemShut
  {NoStop}%
\bibitem [{\citenamefont {Fornieri}\ \emph {et~al.}(2015)\citenamefont
  {Fornieri}, \citenamefont {Mart{\'{i}}nez-P{\'{e}}rez},\ and\ \citenamefont
  {Giazotto}}]{Fornieri2015}%
  \BibitemOpen
  \bibfield  {author} {\bibinfo {author} {\bibfnamefont {A.}~\bibnamefont
  {Fornieri}}, \bibinfo {author} {\bibfnamefont {M.~J.}\ \bibnamefont
  {Mart{\'{i}}nez-P{\'{e}}rez}}, \ and\ \bibinfo {author} {\bibfnamefont
  {F.}~\bibnamefont {Giazotto}},\ }\href
  {https://aip.scitation.org/doi/abs/10.1063/1.4915899} {\bibfield  {journal}
  {\bibinfo  {journal} {AIP Adv.}\ }\textbf {\bibinfo {volume} {5}},\ \bibinfo
  {pages} {053301} (\bibinfo {year} {2015})}\BibitemShut {NoStop}%
\bibitem [{\citenamefont {Giazotto}\ and\ \citenamefont
  {Bergeret}(2020)}]{Giazotto2020}%
  \BibitemOpen
  \bibfield  {author} {\bibinfo {author} {\bibfnamefont {F.}~\bibnamefont
  {Giazotto}}\ and\ \bibinfo {author} {\bibfnamefont {F.~S.}\ \bibnamefont
  {Bergeret}},\ }\href {\doibase 10.1063/5.0010148} {\bibfield  {journal}
  {\bibinfo  {journal} {Appl. Phys. Lett.}\ }\textbf {\bibinfo {volume}
  {116}},\ \bibinfo {pages} {192601} (\bibinfo {year} {2020})}\BibitemShut
  {NoStop}%
\bibitem [{\citenamefont {Marchegiani}\ \emph {et~al.}(2021)\citenamefont
  {Marchegiani}, \citenamefont {Braggio},\ and\ \citenamefont
  {Giazotto}}]{Marchegiani2021}%
  \BibitemOpen
  \bibfield  {author} {\bibinfo {author} {\bibfnamefont {G.}~\bibnamefont
  {Marchegiani}}, \bibinfo {author} {\bibfnamefont {A.}~\bibnamefont
  {Braggio}}, \ and\ \bibinfo {author} {\bibfnamefont {F.}~\bibnamefont
  {Giazotto}},\ }\href {\doibase 10.1063/5.0036485} {\bibfield  {journal}
  {\bibinfo  {journal} {Appl. Phys. Lett.}\ }\textbf {\bibinfo {volume}
  {118}},\ \bibinfo {pages} {022602} (\bibinfo {year} {2021})}\BibitemShut
  {NoStop}%
\bibitem [{\citenamefont {Khomchenko}\ \emph {et~al.}()\citenamefont
  {Khomchenko}, \citenamefont {Ouerdane},\ and\ \citenamefont
  {Benenti}}]{Khomchenko2022}%
  \BibitemOpen
  \bibfield  {author} {\bibinfo {author} {\bibfnamefont {I.}~\bibnamefont
  {Khomchenko}}, \bibinfo {author} {\bibfnamefont {H.}~\bibnamefont
  {Ouerdane}}, \ and\ \bibinfo {author} {\bibfnamefont {G.}~\bibnamefont
  {Benenti}},\ }\href {https://arxiv.org/abs/2206.10600} {\bibinfo  {journal}
  {arXiv:2206.10600 (2022)}\ }\BibitemShut {NoStop}%
\bibitem [{\citenamefont {Ili{\'{c}}}\ \emph {et~al.}(2022)\citenamefont
  {Ili{\'{c}}}, \citenamefont {Virtanen}, \citenamefont {Heikkil\"{a}},\ and\
  \citenamefont {Bergeret}}]{Ili2022}%
  \BibitemOpen
\bibfield  {journal} {  }\bibfield  {author} {\bibinfo {author} {\bibfnamefont
  {S.}~\bibnamefont {Ili{\'{c}}}}, \bibinfo {author} {\bibfnamefont
  {P.}~\bibnamefont {Virtanen}}, \bibinfo {author} {\bibfnamefont {T.~T.}\
  \bibnamefont {Heikkil\"{a}}}, \ and\ \bibinfo {author} {\bibfnamefont
  {F.~S.}\ \bibnamefont {Bergeret}},\ }\href
  {https://doi.org/10.1103/physrevapplied.17.034049} {\bibfield  {journal}
  {\bibinfo  {journal} {Phys. Rev. Appl.}\ }\textbf {\bibinfo {volume} {17}}
  (\bibinfo {year} {2022})}\BibitemShut {NoStop}%
\bibitem [{\citenamefont {Mascarenhas}\ \emph {et~al.}(2014)\citenamefont
  {Mascarenhas}, \citenamefont {Gerace}, \citenamefont {Valente}, \citenamefont
  {Montangero}, \citenamefont {Auff{\`{e}}ves},\ and\ \citenamefont
  {Santos}}]{Mascarenhas2014}%
  \BibitemOpen
  \bibfield  {author} {\bibinfo {author} {\bibfnamefont {E.}~\bibnamefont
  {Mascarenhas}}, \bibinfo {author} {\bibfnamefont {D.}~\bibnamefont {Gerace}},
  \bibinfo {author} {\bibfnamefont {D.}~\bibnamefont {Valente}}, \bibinfo
  {author} {\bibfnamefont {S.}~\bibnamefont {Montangero}}, \bibinfo {author}
  {\bibfnamefont {A.}~\bibnamefont {Auff{\`{e}}ves}}, \ and\ \bibinfo {author}
  {\bibfnamefont {M.~F.}\ \bibnamefont {Santos}},\ }\href
  {https://doi.org/10.1209/0295-5075/106/54003} {\bibfield  {journal} {\bibinfo
   {journal} {{Europhys. Lett.}}\ }\textbf {\bibinfo {volume} {106}},\ \bibinfo
  {pages} {54003} (\bibinfo {year} {2014})}\BibitemShut {NoStop}%
\bibitem [{\citenamefont {Balachandran}\ \emph {et~al.}(2018)\citenamefont
  {Balachandran}, \citenamefont {Benenti}, \citenamefont {Pereira},
  \citenamefont {Casati},\ and\ \citenamefont {Poletti}}]{Balachandran2018bb}%
  \BibitemOpen
  \bibfield  {author} {\bibinfo {author} {\bibfnamefont {V.}~\bibnamefont
  {Balachandran}}, \bibinfo {author} {\bibfnamefont {G.}~\bibnamefont
  {Benenti}}, \bibinfo {author} {\bibfnamefont {E.}~\bibnamefont {Pereira}},
  \bibinfo {author} {\bibfnamefont {G.}~\bibnamefont {Casati}}, \ and\ \bibinfo
  {author} {\bibfnamefont {D.}~\bibnamefont {Poletti}},\ }\href {\doibase
  10.1103/PhysRevLett.120.200603} {\bibfield  {journal} {\bibinfo  {journal}
  {Phys. Rev. Lett.}\ }\textbf {\bibinfo {volume} {120}},\ \bibinfo {pages}
  {200603} (\bibinfo {year} {2018})}\BibitemShut {NoStop}%
\bibitem [{\citenamefont {Lindblad}(1976)}]{Lindblad1976}%
  \BibitemOpen
  \bibfield  {author} {\bibinfo {author} {\bibfnamefont {G.}~\bibnamefont
  {Lindblad}},\ }\href {\doibase 10.1007/BF01608499} {\bibfield  {journal}
  {\bibinfo  {journal} {Comm. Math. Phys.}\ }\textbf {\bibinfo {volume} {48}},\
  \bibinfo {pages} {119} (\bibinfo {year} {1976})}\BibitemShut {NoStop}%
\bibitem [{\citenamefont {Gorini}\ \emph {et~al.}(1976)\citenamefont {Gorini},
  \citenamefont {Kossakowski},\ and\ \citenamefont {Sudarshan}}]{Gorini1976}%
  \BibitemOpen
  \bibfield  {author} {\bibinfo {author} {\bibfnamefont {V.}~\bibnamefont
  {Gorini}}, \bibinfo {author} {\bibfnamefont {A.}~\bibnamefont {Kossakowski}},
  \ and\ \bibinfo {author} {\bibfnamefont {E.~C.}\ \bibnamefont {Sudarshan}},\
  }\href {\doibase 10.1063/1.522979} {\bibfield  {journal} {\bibinfo  {journal}
  {J. Math. Phys.}\ }\textbf {\bibinfo {volume} {17}},\ \bibinfo {pages} {821}
  (\bibinfo {year} {1976})}\BibitemShut {NoStop}%
\bibitem [{\citenamefont {Breuer}\ and\ \citenamefont
  {Petruccione}(2007)}]{Breuer2007}%
  \BibitemOpen
  \bibfield  {author} {\bibinfo {author} {\bibfnamefont {H.-P.}\ \bibnamefont
  {Breuer}}\ and\ \bibinfo {author} {\bibfnamefont {F.}~\bibnamefont
  {Petruccione}},\ }\href {\doibase 10.1093/acprof:oso/9780199213900.001.0001}
  {\emph {\bibinfo {title} {The Theory of Open Quantum Systems}}}\ (\bibinfo
  {publisher} {Oxford University Press},\ \bibinfo {year} {2007})\BibitemShut
  {NoStop}%
\bibitem [{\citenamefont {Hofer}\ \emph {et~al.}(2017)\citenamefont {Hofer},
  \citenamefont {Perarnau-Llobet}, \citenamefont {Miranda}, \citenamefont
  {Haack}, \citenamefont {Silva}, \citenamefont {Brask},\ and\ \citenamefont
  {Brunner}}]{Hofer2017}%
  \BibitemOpen
  \bibfield  {author} {\bibinfo {author} {\bibfnamefont {P.~P.}\ \bibnamefont
  {Hofer}}, \bibinfo {author} {\bibfnamefont {M.}~\bibnamefont
  {Perarnau-Llobet}}, \bibinfo {author} {\bibfnamefont {L.~D.~M.}\ \bibnamefont
  {Miranda}}, \bibinfo {author} {\bibfnamefont {G.}~\bibnamefont {Haack}},
  \bibinfo {author} {\bibfnamefont {R.}~\bibnamefont {Silva}}, \bibinfo
  {author} {\bibfnamefont {J.~B.}\ \bibnamefont {Brask}}, \ and\ \bibinfo
  {author} {\bibfnamefont {N.}~\bibnamefont {Brunner}},\ }\href {\doibase
  10.1088/1367-2630/aa964f} {\bibfield  {journal} {\bibinfo  {journal} {New J.
  Phys.}\ }\textbf {\bibinfo {volume} {19}},\ \bibinfo {pages} {123037}
  (\bibinfo {year} {2017})}\BibitemShut {NoStop}%
\bibitem [{\citenamefont {González}\ \emph {et~al.}(2017)\citenamefont
  {González}, \citenamefont {Correa}, \citenamefont {Nocerino}, \citenamefont
  {Palao}, \citenamefont {Alonso},\ and\ \citenamefont
  {Adesso}}]{Gonzalez2017}%
  \BibitemOpen
  \bibfield  {author} {\bibinfo {author} {\bibfnamefont {J.}~\bibnamefont
  {González}}, \bibinfo {author} {\bibfnamefont {L.}~\bibnamefont {Correa}},
  \bibinfo {author} {\bibfnamefont {G.}~\bibnamefont {Nocerino}}, \bibinfo
  {author} {\bibfnamefont {J.}~\bibnamefont {Palao}}, \bibinfo {author}
  {\bibfnamefont {D.}~\bibnamefont {Alonso}}, \ and\ \bibinfo {author}
  {\bibfnamefont {G.}~\bibnamefont {Adesso}},\ }\href
  {https://doi.org/10.1142/S1230161217400108} {\bibfield  {journal} {\bibinfo
  {journal} {Open Sys. Inf. Dyn.}\ }\textbf {\bibinfo {volume} {24}},\ \bibinfo
  {pages} {1740010} (\bibinfo {year} {2017})}\BibitemShut {NoStop}%
\bibitem [{\citenamefont {Mitchison}\ and\ \citenamefont
  {Plenio}(2018)}]{Mitchison2018}%
  \BibitemOpen
  \bibfield  {author} {\bibinfo {author} {\bibfnamefont {M.~T.}\ \bibnamefont
  {Mitchison}}\ and\ \bibinfo {author} {\bibfnamefont {M.~B.}\ \bibnamefont
  {Plenio}},\ }\href
  {https://iopscience.iop.org/article/10.1088/1367-2630/aa9f70} {\bibfield
  {journal} {\bibinfo  {journal} {New J. Phys.}\ }\textbf {\bibinfo {volume}
  {20}},\ \bibinfo {pages} {033005} (\bibinfo {year} {2018})}\BibitemShut
  {NoStop}%
\bibitem [{\citenamefont {Khandelwal}\ \emph {et~al.}(2020)\citenamefont
  {Khandelwal}, \citenamefont {Palazzo}, \citenamefont {Brunner},\ and\
  \citenamefont {Haack}}]{Khandelwal2020}%
  \BibitemOpen
  \bibfield  {author} {\bibinfo {author} {\bibfnamefont {S.}~\bibnamefont
  {Khandelwal}}, \bibinfo {author} {\bibfnamefont {N.}~\bibnamefont {Palazzo}},
  \bibinfo {author} {\bibfnamefont {N.}~\bibnamefont {Brunner}}, \ and\
  \bibinfo {author} {\bibfnamefont {G.}~\bibnamefont {Haack}},\ }\href
  {\doibase 10.1088/1367-2630/ab9983} {\bibfield  {journal} {\bibinfo
  {journal} {New J. Phys.}\ }\textbf {\bibinfo {volume} {22}},\ \bibinfo
  {pages} {073039} (\bibinfo {year} {2020})}\BibitemShut {NoStop}%
\bibitem [{\citenamefont {Pereira}(2019{\natexlab{a}})}]{Pereira2019}%
  \BibitemOpen
  \bibfield  {author} {\bibinfo {author} {\bibfnamefont {E.}~\bibnamefont
  {Pereira}},\ }\href {https://doi.org/10.1103/physreve.99.032116} {\bibfield
  {journal} {\bibinfo  {journal} {Phys. Rev. E}\ }\textbf {\bibinfo {volume}
  {99}} (\bibinfo {year} {2019}{\natexlab{a}})}\BibitemShut {NoStop}%
\bibitem [{\citenamefont {Pereira}(2019{\natexlab{b}})}]{Pereira2019II}%
  \BibitemOpen
  \bibfield  {author} {\bibinfo {author} {\bibfnamefont {E.}~\bibnamefont
  {Pereira}},\ }\href {https://doi.org/10.1209/0295-5075/126/14001} {\bibfield
  {journal} {\bibinfo  {journal} {Eur. Phys. Lett.}\ }\textbf {\bibinfo
  {volume} {126}},\ \bibinfo {pages} {14001} (\bibinfo {year}
  {2019}{\natexlab{b}})}\BibitemShut {NoStop}%
\end{thebibliography}%


%

\end{document}